\documentclass{style}
\usepackage{graphicx}% Include figure files
\usepackage{dcolumn}% Align table columns on decimal point
\usepackage{bm}% bold math
 \def\ba#1{\begin{array}{#1}}
 \def\ea{\end{array}}
 \def\be{\begin{equation}}
 \def\ee{\end{equation}}
\def\bq{\begin{equation}}
\def\eq{\end{equation}}
 \def\br{\begin{eqnarray}}
 \def\er{\end{eqnarray}}

 \def\sp#1{\langle #1 \rangle }

\begin{document}
\title{Introduction to low-momentum effective interactions with \\
Brown-Rho scaling and three-nucleon forces }

\author{
T. T. S. Kuo \\
\it Department of Physics and Astronomy, Stony Brook University, 
Stony Brook, NY 11794, USA,\\ \\
J. W. Holt \\
\it Department of Physics, University of Washington, 
Seattle, WA 98195, USA, \\ \\
E. Osnes \\
\it Institute of Physics, University of Oslo, NO-0316 Oslo, Norway
}

\pacs{}
\maketitle

%\date{\today}

\begin{abstract}
Model-space effective interactions $V_{eff}$ derived from free-space nucleon-nucleon 
interactions $V_{NN}$ are reviewed. We employ a double decimation approach: first we 
extract a low-momentum interaction $V_{low-k}$ from $V_{NN}$ using a $T$-matrix 
equivalence decimation method. Then $V_{eff}$ is obtained from $V_{low-k}$ by way of 
a folded-diagram effective interaction method. For decimation momentum $\Lambda \simeq 
2 fm^{-1}$, the $V_{low-k}$ interactions derived from different realistic $V_{NN}$ models 
are nearly model independent, and so are the resulting shell-model effective interactions. 
For nucleons in a low-density nuclear medium like valence nucleons near the nuclear surface, 
such effective interactions derived from free-space $V_{NN}$ are satisfactory in reproducing 
experimental nuclear properties. But it is not so for nucleons in a nuclear medium with density 
near or beyond nuclear matter saturation density. In this case it may be necessary to include 
the effects from Brown-Rho (BR) scaling of hadrons and/or three-nucleon forces $V_{3N}$, 
effectively changing the free-space $V_{NN}$ into a density-dependent one. The 
density-dependent effects from BR scaling and $V_{3N}$ are compared with those from 
empirical Skyrme effective interactions.

\end{abstract}

\section{Introduction  }
\vskip 0.5cm
In treating nuclear many-body problems, one can often reduce
the ``full" many-body
problem to a much smaller and more manageable problem, referred to
as a model-space problem. In so doing, an important step is
to determine the  model-space 
effective interaction $V_{eff}$. 
In this paper, we would like to present an introductory and 
pedagogical review of this topic, especially the
derivation of $V_{eff}$ from  realistic 
meson-exchange nucleon-nucleon interactions.

 Many-body problems are difficult  
as we are all aware of.
 They are particularly so, when A, the number of 
particles in the system, is large. In fact a many-body problem with 
A=3 is already a very hard problem. In the real world, A is usually 
much larger. For example, A=18 when the nucleus $^{18}O$
is treated as an 18-nucleon problem. It becomes an A=54 problem
if it is treated as composed of 54 quarks. So the complexity of the
problem depends largely on how we look at the problem,
or, more precisely, 
it depends on what is our ``model space" within which we are making our
observations.

Let us consider the nucleus $^{18}O$ as an example;  it is a system of
18 nucleons. The nuclear shell model has been the most successful
model for nuclear structure. This model is in fact a model-space
approach. The wave functions $\Psi _n$ of this nucleus consists of, 
in principle, shell-model wave functions 
$|2p0h\rangle$, 
$|3p1h\rangle$, .... and 
$|18p16h\rangle$; $p$ and $h$ denoting particles and holes
with respect to a closed $^{16}O$ core.
 The dimension of this full shell-model space is practically
``infinite". For actual calculations, some truncation or renormalization
of the full-space problem is indispensible.
To describe the low-energy properties of this nucleus, it may not be necessary
to consider all these wave-function components; it may be sufficient
to include just some low-energy parts of them.
This is actually the approach commonly used in  the  shell-model description
of $^{18}O$, where this nucleus is treated as composed of
 two valence neutrons in the 0d1s shell outside a closed $^{16}O$ core. 
Denoting this restricted model space as P, the Hamiltonian for this
nucleus is $P(H_0+V_{eff})P$ 
where  $V_{eff}$
is the effective interaction and $H_0$ denotes the single-particle (sp)
Hamiltonian.

\begin{table}[hb]
\begin{center}
\begin{tabular}{lll}
\hline \hline
Nucleus & Experimental & Calculated  \\
\hline
$^{41}Ca$ & 8.36 &8.38 \\
$^{42}Ca$ & 19.83 &19.86 \\
$^{43}Ca$ & 27.75 &27.78 \\
$^{44}Ca$ & 38.89 &38.80 \\
$^{45}Ca$ & 46.31 &46.26 \\
$^{46}Ca$ & 56.72 &56.82 \\
$^{47}Ca$ &  &63.81 \\
$^{48}Ca$ & 73.95 &73.93 \\
\hline
\end{tabular}
\caption{Binding energies of Ca isotopes in MeV
 given by Talmi's $(0f_{7/2})^n$ model.\cite{talmi}}
\end{center}
\end{table}

  In such  model-space approaches, an important step is to employ
 some empirically determined  effective interactions
$V_{eff}$. And indeed  such approaches have  been  very successful. 
To illustrate this, let us  consider 
the pioneering work of Talmi \cite{talmi} on
the calcium isotopes $^{41}Ca - {^{48}Ca}$. In this work these isotopes
 were treated as composed of $n$ valence neutrons
in the $0f_{7/2}$ shell outside an inert $^{40}Ca$ core. In other words,
a restricted model space of $(0f_{7/2})^n$ is employed.
For this model space, $V_{eff}$ can be expressed in terms of three parameters.
 In fact the ground-state energies 
are given by a very simple formula 
\be
\langle (0f_{7/2})^n|H_{eff}|(0f_{7/2})^n\rangle=nC+\frac{n(n-1)}{2}\alpha +[\frac{1}{2}n]\beta
\ee
where $[\frac{1}{2}n]$ is the step function equal to $\frac{n}{2}$
if $n$=even, and $\frac{n-1}{2}$ if $n$= odd.
The parameters $C$, $\alpha$ and $\beta$ can be determined by fitting
the experimental energies. The optimum values, in MeV,
 are determined as
$$C=8.38\pm0.05,~~
\alpha=-0.21\pm0.01,~~
\beta=3.33\pm0.12.$$

The binding energies given by these parameters are compared with experiments
in Table 1. As seen the agreement is astonishingly excellent. The term $C$
is just the neutron sp energy which can be extracted from the experimental
binding energies of $^{41}Ca$ and $^{40}Ca$. We note that the optimum
value of $C$ is practically the same as the experimental sp energy
(8.38 vs 8.36).  $C$ is actually not a parameter; it is the experimental
sp energy. Thus in this model space approach, there are only two parameters,
$\alpha$ and $\beta$, which characterize the effective interaction $V_{eff}$.
The above example strongly indicates the success of
 the model-space approach: by confining
the nucleons to a small (manageable) space and treating $V_{eff}$ as composed
of adjustable parameters, experimental results can indeed 
be quite satisfactorily
reproduced.
\begin{figure}
\begin{center}
%\scalebox{0.34}{\includegraphics[angle=-90]{715vbare1s01s0.eps5v}}
\scalebox{0.32}{\includegraphics[angle=-90]{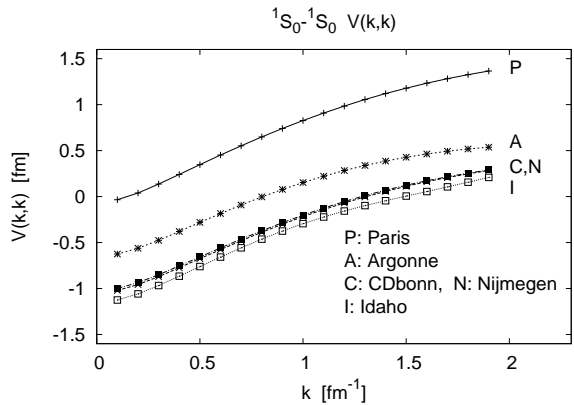}}
%figureeps
\caption{  Comparison of the $^1S_0 - {^1S_0}$ diagonal 
momentum-space matrix elements
$V(k,k)$ of Paris \cite{paris}, 
CD-Bonn \cite{cdbonn}, 
Argonne \cite{argonne}, 
Nijmegen \cite{nijmegen} 
and Idaho \cite{chiralvnn} potentials.}
\end{center}
\end{figure}

\begin{figure}
\begin{center}
%\scalebox{0.34}{\includegraphics[angle=-90]{715vbare3s13s1.eps5v}}
\scalebox{0.32}{\includegraphics[angle=-90]{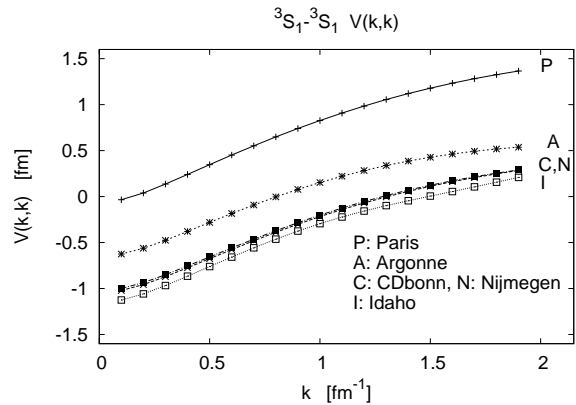}}
%figureeps
\caption{Same as Fig.\ 1 except for  the $^3S_1-{^3S_1}$ channel.}
\end{center}
\end{figure}

 A challenging task is to see if we can 
derive microscopically the empirical
$V_{eff}$ from an underlying nucleon-nucleon (NN) interaction.
%(In the present work we shall use the notations
% $V_{NN}$, $V_{3N}$, $\cdots$ to denote 
%respectively the two-nucleon, three-nucleon, $\cdots$ nucleon-nucleon
%interaction.)
Since the early works of Brown and Kuo \cite{brownkuo67,kuobrown66},
this question has been rather extensively studied (see e.g.\ Refs.\
\cite{jensen95,corag09} and references quoted therein).
In microscopic nuclear structure calculations starting from
 $V_{NN}$, a well-known ambiguity
has been  the choice of  NN potential. There are a number
of successful models for $V_{NN}$, such as the Paris \cite{paris},
Bonn \cite{mach89}, CD-Bonn \cite{cdbonn},
Argonne V18 \cite{argonne}, Nijmegen \cite{nijmegen}
and the chiral Idaho \cite{chiralvnn} potentials. A common feature of these
potentials is that they all  
reproduce  the empirical deuteron properties and low-energy
phase shifts very accurately. But, as illustrated in Figs.\ 1 and 2, 
 these potentials are in fact significantly different from each other
in the momentum representation.
This has been a situation of much concern. Certainly we would like
to have a ``unique" NN potential (like the Coulomb potential
between two electric charges). As seen, the above potentials are not
`unique'. Which of them is the `right' NN interaction?
We shall make an effort  addressing this question.

   The present paper is organized as follows. In section II
we shall describe in some detail  a $\hat Q$-box folded-diagram
expansion for the  model-space effective interaction 
$V_{eff}$ for valence nuclei such as $^{18}O$ and $^{19}O$
confined within a model space $P$. 
In this formalism the effective Hamiltonian
is of the form $PH_{eff}P=P(H_0^{expt}+V_{eff})P$ where
 the sp Hamiltonian
$H_0^{expt}$ is obtained from   the experimental energies
of nuclei with one valence nucleon such as $^{17}O$.
 The general structure of this
formalism is compared with the Talmi approach described earlier.
 Methods for summing up the
folded-diagram series for degenerate and non-degenerate,
such as  two-major-shell,  model
spaces are discussed, and with them
the present framework can be used 
to calculate the model-space $V_{eff}$
 from input nucleon-nucleon interactions.

  In section III we shall review a $T$-matrix equivalence 
approach for deriving the low-momentum NN interaction 
$V_{low-k}$ by integrating, or decimating, out the 
high-momentum modes of 
the input interaction. Momentum space matrix elements
of $V_{low-k}$ calculated from different $V_{NN}$ potentials
are compared.  The counter terms generated by the above  decimation
are discussed. The $V_{low-k}$ interaction so constructed is non-Hermitian
and transformation methods for making it Hermitian are described.
In section IV we shall describe Brown-Rho (BR) scaling, 
three-nucleon forces, and the density-dependent effects
generated by them. Applications  to nuclear structure,
nuclear matter and neutron stars will be reported.
A summary and discussion will be presented in section V.

\section{Folded-diagram expansion for effective interactions $V_{eff}$}

In this section we shall describe some details of the 
folded-diagram expansion for the shell-model
effective interaction $V_{eff}$ \cite{kuos,klr71}.
Let us first discuss a similarity between this expansion
and that for the ground-state energy of a closed-shell (or filled Fermi-sea)
system such as  $^{16}O$ (or nuclear matter). 
Consider the nuclear matter case, whose
true and unperturbed ground-state energies are   denoted by
$E_0$ and $W_0$ respectively. The  ground-state
energy shift $\Delta E_0=E_0-W_0$ is given by the
 Goldstone linked-diagram expansion
\cite{kuos,klr71,goldstone}, namely  the sum of all the 
 linked diagrams, as shown in Fig.\ 3.
Here diagram (a) is the lowest order such diagram (usually referred
to as the Hartree-Fock interaction diagram). Diagram (b) is
a particle-particle ladder diagram  having repeated interactions
between a pair of particle lines. Diagram (c) is a ring diagram
(see e.g., Ref.\ \cite{siu09,dong09}) with both particle-particle
and hole-hole interactions.
\begin{figure}[here]
\begin{center}
\scalebox{0.4}{\includegraphics{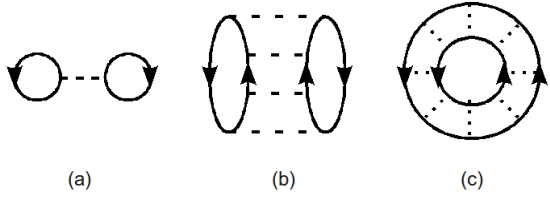}}
\caption{Sample  diagrams contained in the Goldstone linked
diagram expansion for the ground-state energy shift $\Delta E_0$.
Each dashed line represents a nuclear interaction vertex.}\label{figure1}
\end{center}
\end{figure}
\begin{figure}
\begin{center}
\scalebox{0.3}{
\includegraphics[angle=-90]{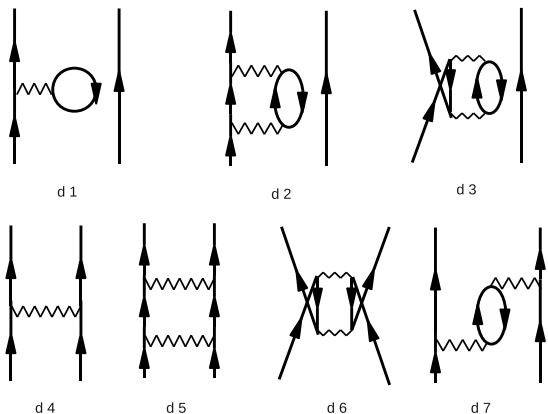}
}
\caption{Low-order diagrams belonging to the $\hat Q$-box for $^{18}O$.
Note that the intermediate states between two successive vertices
must be outside the chosen model space as discussed later.}
\end{center}
\end{figure}

 The  Goldstone expansion provides a framework
for calculating, for example, the ground-state energy of $^{16}O$.
What would then be the corresponding framework for calculating 
the low-lying $0^+$ states of $^{18}O$?
It is here the folded-diagram expansion comes in.
This  expansion  provides  a framework for 
such and similar calculations. 
It is a generalization of 
the Goldstone expansion 
to  systems with valence particles confined to a multi-dimensional
model space.
 Consider again $^{18}O$ as an example. 
This nucleus has two valence neutrons,
and as shown in Fig.\ 4 the diagrams of its effective interaction
all have two valence lines in their initial and final states.
 (Note that the initial and final states
of the Goldstone diagrams as shown in Fig.\ 3 are both vacuum.)
In a shell-model calculation
of the low-lying $0^+$ states of $^{18}O$, one usually treats it as 
two valence neutrons confined  in the $0d1s$ shell with 
an effective Hamiltonian
$H_{eff}=H_0+V_{eff}$. In this case the $V_{eff}$ for the valence
neutrons is a matrix of dimension three. One  obtains both the ground-
 and excited-state  energies
of the system
by solving a matrix equation involving $V_{eff}$. (In the Goldstone
expansion, the ground-state energy is given by the sum of
all the linked diagrams.)
The purpose of the folded-diagram expansion \cite{kuos,klr71} 
is to  provide
 a microscopic framework
to derive such $V_{eff}$ from
an underlying $V_{NN}$ potential.

The nuclear many-body
Schroedinger equation may be written as 
\br
H\Psi_{n}(1,2,..A)&=&E_{n}(1,2,..A)\Psi_{n}(1,2,...A);
\nonumber \\
 H&=& T+V_{NN}
\er
where $T$ denotes the kinetic energy. To define a convenient
sp basis, we introduce an auxiliary potential $U$ and rewrite
$H$ as
\bq
H=H_0 + H_1; ~H_0=T+U,~H_1=V_{NN}-U.
\eq
The choice of $U$ is very important, and in principle we 
can use any $U$ of our
choice. However, an optimized choice is to have a $U$
 such that $H_1$  becomes 
``small". In this way $H_1$ may  be treated as a perturbation. 
 We denote the sp wave functions 
and energies defined by $H_0$ as
$\phi$ and $\epsilon$, namely $H_0 \phi _n =  \epsilon _n \phi _n$.

The many-body problem as specified by Eq.\ (2) is in general very difficult.
It has a  large number of 
solutions. In fact we may not need to know all of them; perhaps only a few
 of them are of physical interest and we should just calculate these few
solutions. Thus we aim at the reduction of Eq.\ (2) to
 a model-space equation of the form
\bq
PH_{eff}P\Psi_{m}=E_{m}P\Psi_{m};~m=1,\cdots ,d
\eq
where $P$ is the model-space projection operator defined by
\bq
P=\sum_{ n=1,d} \vert \Phi_n \rangle \langle \Phi _n \vert.
\eq
Here $\Phi$ is a Slater
determinant composed of the sp
 wave functions $\phi$. The above equation
reproduces only $d$ solutions of Eq.\ (2). Furthermore, it does 
not give the complete
wave function. It gives only the projection of the whole wave function
onto the model space ($P$) of one's choice. 

The main idea behind Eq.\ (4) is to have a smaller, and hopefully more 
manageable, many-body problem than the original problem of Eq.\ (2).
This approach is useful if the effective Hamiltonian can be obtained
without too much difficulty. Hence  the question now is how to derive 
 $H_{eff}$, or how to derive the effective interaction $V_{eff}$
which is related to $H_{eff}$ by
\bq
H_{eff}=H_0+V_{eff}.
\eq

There are various ways to obtain $H_{eff}$ or $V_{eff}$. Let us first describe
briefly the Feshbach \cite{fesh} formulation of the model-space 
effective Hamiltonian. In matrix form Eq.\ (2) is written as
\br
\left( \begin{array}{cc}
 P H P & P H Q \\
 Q H P & Q H Q 
\end{array} \right)
\left( \begin{array}{c}
P\Psi \\
Q\Psi 
\end{array} \right)
= E \left( \begin{array}{c}
P\Psi \\
Q\Psi 
\end{array} \right)
\er
where $Q$ is the complement of $P$, namely $Q=1-P$. We can readily eliminate
$Q\Psi$, obtaining a $P$-space equation
\bq
H_{eff}(E_n) P\Psi_{n}= E_n P\Psi_{n}
\eq
with
\bq
H_{eff}(E_n)= PHP +  PHQ\frac{1}{E_n -QHQ}QHP.
\eq
This is an interesting result. We see that now we only need to deal 
with an equation which is entirely contained in the $P$ space. However,  
 a significant drawback here is 
 that the effective Hamiltonian is dependent on the eigenvalue $E_n$.
In other words, we need to use different effective 
interactions for different
eigenstates. This feature is clearly not convenient or desirable. 

 The  folded-diagram method has been developed with the purpose of 
providing an energy-independent (namely independent of the
energy eigenvalue $E_n$)
model-space effective interaction.
There have been several studies of the folded-diagram method
\cite{kuos,klr71,morita,brandow,johnson,lindgren}. 
Among them, the time-dependent
formulation of Kuo, Lee and Ratcliff (KLR) \cite{kuos,klr71} is particularly 
suitable for  shell-model calculations. It provides a formal framework
for reducing the full-space many-body problem, as shown by Eq.\ (2), to a 
model-space one of the form
\br
PH_{eff}P\Psi_m&=&(E_m-E_0^{core})P\Psi_m; \nonumber \\
PH_{eff}P&=&P[H_0^{expt}+V_{eff}]P
\er
where $m=1,...,d$; $d$ being the dimension of the model space.
 Note that this 
equation calculates the energy difference between
 neighboring nuclei. For example,
$E_m$ is the energy of $^{19}O$ and $E_0^{core}$ is the ground-state
 energy of $^{16}O$. In addition, $H_0^{expt}$ is the sp energy part
which is extracted from experimental energy differences between neighboring
nuclei such as $^{17}O$ and $^{16}O$.
The above $H_{eff}$ is of the same form as the  
one commonly used in shell-model calculations (such as the calculation
of the $Ca$ isotopes mentioned earlier), except 
that here $V_{eff}$ is  derived
from the nucleon-nucleon interaction $V_{NN}$ whereas in empirical
shell-model calculations it is determined empirically by fitting
experimental data. 
The folded-diagram method has been applied to a wide range of
nuclear structure calculations using realistic NN
interactions \cite{jensen95,corag09,holtc1408}. 

%from kuo-osnes p.35
\vskip 0.3cm
In deriving $V_{eff}$ microscopically, we shall employ 
the time evolution operator
$U(t,t')=exp[-iH(t-t')]$
 in the complex-time limit, namely 
\begin{equation}
\lim_{t'\rightarrow -\infty(\epsilon)}\equiv 
 \lim_{\epsilon\rightarrow 0^+} 
 \lim_{t'\rightarrow -\infty(1-i\epsilon)}
\end{equation}
In this limit it can be  shown that we can construct, starting from
a model-space parent
state $\rho _i$, the lowest eigenstate $\Psi _i$ of the true
Hamiltonian with $\langle \rho _i | \Psi _i   \rangle \neq 0$, namely
\begin{equation}
\lim _{t'\rightarrow -\infty (\epsilon)}
\frac{U(0,t')|\rho _i\rangle}{\langle \rho_i|U(0,t')|\rho_i \rangle}
=\frac{|\Psi _i\rangle}{\langle \rho_i|\Psi _i\rangle},
\end{equation}
where the parent state is a linear combination of the model-space
basis vectors $\Phi _k$,
\begin{equation}
|\rho _{\lambda}\rangle =\sum _{k=1,d}C_{\lambda,k}|\Phi _k\rangle ,
\end{equation}
 $d$ being the dimension of the model space.
When $P\Psi _{\lambda}$ for $\lambda=1,..., d$ 
are linearly 
independent, the $C_{\lambda,k}$ coefficients can satisfy
\be
\langle \rho _{\lambda} |\Psi _{\mu}\rangle
=\langle \rho _{\lambda} |P\Psi _{\mu}\rangle=0,
\ee
for $\lambda \ne \mu; ~\lambda,\mu=1,..., d$.
Then we have 
\be
H
\frac{U(0,t')|\rho _{\lambda}\rangle}{\langle \rho_{\lambda}|U(0,t')|\rho_{\lambda}\rangle} =
E_{\lambda} \frac{U(0,t')|\rho _{\lambda}\rangle}
{\langle \rho_{\lambda}|U(0,t')|\rho_{\lambda}\rangle}, 
\ee
where $\lambda=1,..., d$ with the complex-time limit
$t'\rightarrow -\infty (\epsilon)$ understood.
\vskip 0.5cm

 Eq.\ (15) is the basic equation for deriving the model-space
effective interaction $V_{eff}$ using a folded-diagram
factorization procedure, which has been given in detail
in \cite{kuos,klr71}. Here we shall just outline some basic
features of the procedure
such as ``what is a folded diagram".

In the interaction representation, we have
\begin{eqnarray}
U(t, t^{'}) &=& 1 + \sum_{n=1}^{\infty} (-i)^{n}\int_{t^{'}}^{t}dt_{1}
\int_{t^{'}}^{t_{1}}dt_{2}\cdots \nonumber\\
&&\int_{t^{'}}^{t_{n-1}}dt_{n}H_{1}(t_{1})H_{1}(t_{2})\cdots H_{1}(t_{n}).
\end{eqnarray}
Consider the wave function generated by the operation $U(0,-\infty)\vert
\Phi_{\alpha \beta} \rangle$ where 
$\vert \Phi_{\alpha \beta} \rangle$ is a two-particle state defined by
$a^{+}_{\alpha}a^{+}_{\beta}\vert 0 \rangle$. Here the $a^+$'s are the sp
creation operators and $\vert 0 \rangle$ is the vacuum state. 
 As indicated in Fig.\ 5,
diagram (A) is a term generated in this operation, where particles
$\alpha, \beta$ have one interaction at time $t_2$, go into intermediate states
$\gamma, \delta$, have one more interaction at $t_1$, and finally end up in 
state $\Phi_{ij}$. We use the notation that``railed" fermion lines denote
 passive sp states, namely those outside the model space $P$, and the bare
fermion lines denote active sp states which are inside $P$. Thus $\alpha,\beta,
\gamma,\delta$ of the figure are within $P$ while $i$ and $j$ belong to $Q$.

%\vskip 7cm
\begin{figure}
\begin{center}
%\centerline{\includegraphics[scale=0.4]{dongfig1.ps}}
\centerline{\includegraphics[scale=0.4]{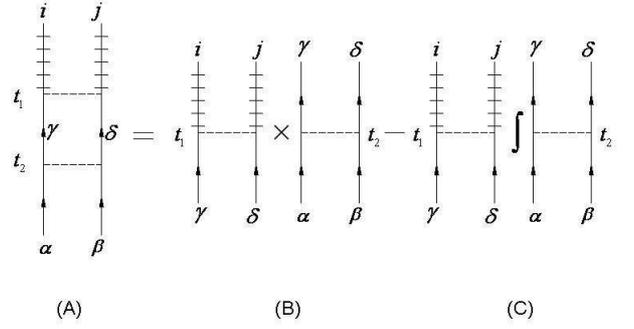}}
\caption{  An example of folded-diagram factorization.}
\end{center}
\end{figure}

The concept of folding can be illustrated by the following factorization
operation. The time integral contained in diagram (A) has the limits
$0>t_1>t_2>-\infty$, i.e. $\int^0_{-\infty}dt_1\int^{t_1}_{-\infty}dt_2$.
For (B), the integrations over $t_1$ and $t_2$ are independent, both
from $-\infty$ to 0. Clearly (A) is not equal to (B). The folded
diagram in this case is defined as the ``error" introduced by
 the factorization
of (A) into (B), namely diagram (C) is the folded diagram given by
\bq
(A)=(B)-(C).
\eq
In fact they have the values
\bq
(A) = \vert \Phi_{ij} \rangle
\frac{V_{ij,\gamma\delta} V_{\gamma\delta, \alpha\beta}}
{( \epsilon_{\alpha} + \epsilon_{\beta} -\epsilon_{i} - \epsilon_{j})
 (\epsilon_{\alpha} + \epsilon_{\beta} -\epsilon_{\gamma} - 
\epsilon_{\delta})},
\eq

\bq
(B) =  \vert \Phi_{ij} \rangle
\frac{V_{ij,\gamma\delta} V_{\gamma\delta, \alpha\beta}}
{( \epsilon_{\gamma} + \epsilon_{\delta} -\epsilon_{i} - \epsilon_{j})
 (\epsilon_{\alpha} + \epsilon_{\beta} -\epsilon_{\gamma} - 
\epsilon_{\delta})},
\eq

\bq
(C) =  \vert \Phi_{ij} \rangle
\frac{V_{ij,\gamma\delta} V_{\gamma\delta, \alpha\beta}}
{( \epsilon_{\gamma} + \epsilon_{\delta} -\epsilon_{i} - \epsilon_{j})
 (\epsilon_{\alpha} + \epsilon_{\beta} -\epsilon_{i} - \epsilon_{j})}.
\eq
When we have a degenerate model space, then $(\epsilon_{\alpha} 
+ \epsilon_{\beta}) =(\epsilon_{\gamma} + \epsilon_{\delta})$ and (A)
and (B) both become divergent. It is of interest 
that  the folded diagram (C)
is still well defined in this case; (C) is a finite quantity
extracted from two divergent ones.

 Using the above folded-diagram factorization, one can obtain an 
energy-independent effective interaction given by \cite{kuos,klr71}
\br
V_{eff}&=& \hat{Q} - \hat{Q}^{'}\int\hat{Q}
+ \hat{Q}^{'}\int\hat{Q}\int\hat{Q} \nonumber \\
&& - \hat{Q}^{'}\int\hat{Q}\int\hat{Q}
\int\hat{Q}\cdots.
\er
This is a well-known folded-diagram expansion for the model-space
effective interaction \cite{kuos,klr71}, which is energy-independent 
and  valence-linked.
For the case of a one-dimensional model space, the above expansion
reduces to the well-known Goldstone linked-diagram expansion
\cite{kuos,goldstone}. The folded-diagram expansion is basically
an extension of the Goldstone expansion to a 
multi-dimensional model space.

Let us now explain the various terms of Eq.\ (21). 
 Each integral sign in the above denotes a ``fold" \cite{klr71}. 
For instance the last term in the equation is a three-fold term; it has three
 integral signs. $\hat Q$ represents a so-called $\hat Q$-box, which may be
schematically written as
\bq
\hat Q(\omega)=[PVP+PVQ\frac{1}{\omega -QHQ}QVP]_{linked}.
\eq
(For convenience, we shall use $V$ in place of $V_{NN}$.)
In fact the $\hat Q$-box is an 
irreducible vertex function where the intermediate 
states between any two vertices must belong to the $Q$ space. 
(Note that we use $Q$, without hat, to denote the $Q$-space 
projection operator.)
It contains valence-linked 
diagrams only, as indicated by the subscript ``linked''. The $\hat Q'$-box of
Eq.\ (21) is defined as ($\hat Q -PVP$). In Fig.\ 4 some low-order
$\hat Q$-box diagrams for nuclei with two valence nucleons, such
as $^{18}O$, are shown. Note that the intermediate state between two
successive vertices must be in the $Q$-space. For example, in an sd-shell 
calculation of $^{18}O$ the intermediate state of diagram
d5 of Fig.\ 4 must have at least one particle outside the sd-shell.

 For a nucleus of $n$ valence nucleons, such as $n=3$ for $^{19}O$,
the above $V_{eff}$ contains one-body (1b), two-body (2b),...and up to
$n$-body ($n$b) terms \cite{kuos}, namely
\br
V_{eff}&=&
V_{eff}(1b)
+V_{eff}(2b)
+V_{eff}(3b)+\cdots \nonumber \\
&& +V_{eff}(nb).
\er
An important feature of this $V_{eff}$ is its ``nucleus independence".
As an example, the $V_{eff}(2b)$ for nuclei $^{A}O$, $A=18,19,20,...$ 
are all the same
according to the above folded-diagram method. This is a desirable feature,
as it allows us to use the same two-body effective 
interaction derived for $A=18$
for all the other sd-shell nuclei. However the many-body forces for
different nuclei are different. For example $^{19}O$ and $^{20}O$ 
have the same $V_{eff}(3b)$, but $^{20}O$ has in addition 
$V_{eff}(4b)$ while $^{19}O$ does not.

 The above $V_{eff}$ allows us to use the experimental sp energies.
To illustrate, let us consider the 0d1s-shell nuclei. The energies
of $^{17}O$ is given by $(H_0+V_{eff}(1b))$ which corresponds to the
experimental sp energies of $^{17}O$. Since $V_{eff}(1b)$ is the
same for all the 0d1s-shell nuclei, we can use these sp energies 
for all of them. Thus the model-space secular equation is of the form
shown in Eq.\ (10). Note, however, when experimental sp energies are employed,
the effective interaction is at least two-body, namely
$H_{eff}=H_0^{expt}+V_{eff}(nb);n\geq 2$.
As indicated in Eq.\ (10), the energies given by $H_{eff}$ are $E_m-E_0^{core}$.
This subtraction is because
the diagrams contained in the folded-diagram expansion are all
valenced-linked, unlinked core-excitation diagrams having been
removed \cite{kuos,klr71}.

We now come to the calculation of the folded-diagram expansion for the
effective interaction.
 The expansion of Eq.\ (21) may be rewritten as
\bq
V_{eff} = F_{0} + F_{1} + F_{2} + F_{3} + F_{4} + \cdots,
\eq
where $F_n$ denotes all the diagrams with $n$ folds. 
If the model space is degenerate, i.e.\ $PH_0P=W_0$ is constant, 
then the various $F$ terms can be conveniently written as
\br
F_0 &=& \hat{Q} \nonumber \\ 
F_1 &=& \hat{Q_1}\hat{Q} \nonumber \\
F_2 &=& \hat{Q_2}\hat{Q}\hat{Q} + \hat{Q_1}\hat{Q} \nonumber \\
F_3 &=& \hat{Q_3}\hat{Q}\hat{Q}\hat{Q}  
 + \hat{Q_2}\hat{Q_1}\hat{Q}\hat{Q} +\hat{Q_2}\hat{Q}\hat{Q_1}\hat{Q}
 \nonumber \\
&& +\hat{Q_1}\hat{Q_2}\hat{Q}\hat{Q} +\hat{Q_1}\hat{Q_1}\hat{Q_1}\hat{Q}
 \nonumber \\
\vdots
\er
where the $\hat Q$-box derivatives are 
\bq
\hat{Q}_{n} = \left . \frac{1}{n!}\frac{d^{n}\hat{Q}}{d\omega^{n}}\right |_{\omega =
W_{0}}.
\eq
In an early calculation for the 0d1s shell \cite{shurpin83}, the above
 folded-diagram series was found to converge 
rather rapidly; the main contribution
was from the 
once-folded term $F_1$, $F_2$ was much smaller and $F_3$ was negligible.
However, it would be useful if the series can be summed to all orders.

 Two iteration methods for numerically calculating the 
folded-diagram series to all orders have been developed. One is  the 
 Krenciglowa-Kuo (KK) iteration method \cite{kren}. 
This method employs a partial summation framework which 
has a clear diagrammatic struture, explicitly showing
how the folded diagrams are generated and summed 
in each additional iteration.
Another iteration method is the Lee-Suzuki method \cite{lesu,sule},
which has been widely employed in shell-model calculations
\cite{jensen95,corag09}. 

Let us outline the Lee-Suzuki method. We define the effective interaction
given by the $n$th iteration as $R_n$, and the initial condition
is chosen as
\bq
R_1=\hat Q(\omega=W_0),
\eq 
where $\hat Q$ is the irreducible vertex function of Eq.\ (22). We consider
a degenerate model space with $PH_0P=W_0$. The effective interaction
for the second iteration is 
\bq
R_{2}= \frac{1}{1- \frac{d\hat{Q}}{d\omega}\mid_{\omega=W_0}}
\times\left[ H_0 -W_0 + \hat{Q}(W_0)\right],
\eq
and for the $n$th iteration we have
\br
R_{n} &=& \frac{1}{1-\hat{Q}_1 -\sum_{m=2}^{n-1} \hat{Q}_m \prod_{k=n-m+1}
^{n-1} R_{k} }
 \nonumber \\
&&\times \left[ H_0 -W_0 +\hat{Q}(W_0) \right],
\er
where $\hat{Q}_m$ is the energy derivative of the $\hat Q$-box as given
by Eq.\ (26). The Lee-Suzuki method is convenient for calculations,
and it usually converges after 3 or 4 iterations.

  A difference between the Krenciglowa-Kuo and Lee-Suzuki iteration methods may be pointed
out. Recall that for a model space of dimension $d$, $PH_{eff}P$ reproduces
only $d$ eigenstates of the full-space $H$. The question is then which $d$
states of $H$ are reproduced by $H_{eff}$. It is of interest that
the Lee-Suzuki method reproduces the $d$ lowest (in energy) states
of $H$ \cite{lesu,sule} while the states reproduced by the 
Krenciglowa-Kuo method are those with the largest $P$-space overlaps \cite{kren}.

  The above methods were originally both formulated for the case
of a degenerate $P$-space, namely $PH_0P$ is degenerate. 
Iteration methods for calculating the effective interaction  
 with non-degenerate
$PH_0P$ have also been developed \cite{suzuki94,kuo95,otsuka14,dong14}.
The method described in \cite{kuo95} is a simple
iteration method, which is outlined below.
We denote the effective interaction for the {\it i}th iteration as
$V_{eff}^{(i)}$ with the corresponding eigenvalues $E$ and eigenfunctions
$\chi$  given by
\begin{equation}
[PH_0P +V_{eff}^{(i)}]\chi_m^{(i)}=E_m^{(i)} \chi_m^{(i)}.
\end{equation}
Here  $\chi_m$ is the $P$-space projection of the full-space eigenfunction
$\Psi_m$, namely $\chi _m= P\Psi _m$.
The effective interaction for the next iteration is then
\br
V_{eff}^{(i+1)}&=&\sum_m[PH_0P+\hat Q(E_m^{(i)})]
| \chi_m^{(i)}\rangle \langle \tilde{\chi} _m^{(i)}| \nonumber \\
&& -PH_0P,
\label{e210}
\er
where the bi-orthogonal states are defined by
\begin{equation}
 \langle \tilde \chi _m |\chi _{m'} \rangle = \delta _{m,m'}. 
\end{equation}
Note that in the above $PH_0P$ is non-degenerate. The converged eigenvalue
$E_m$ and eigenfunction $\chi _m$ satisfy the $P$-space 
self-consistent condition
\bq
(E_m(\omega)-H_0)\chi_m=\hat Q(\omega)\chi _m; \, \, \omega=E_m(\omega).
\label{e212}
\eq 
To start the iteration, we may use
\begin{equation}
V_{eff}^{(1)}=\hat Q(\omega _0),
\end{equation}
where $\omega _0$ is a starting energy chosen to be close to $PH_0P$.
 The converged  effective interaction is given by 
$V_{eff}=V_{eff}^{(n+1)}=V_{eff}^{(n)}$. 
When convergent, the  resultant $V_{eff}$  is independent
of $\omega _0$, as  it is the states
with maximum $P$-space overlaps which are selected 
by this method \cite{kren,kuo95}. 

%For the effective interaction with non-degenerate $PH_0P$, 
%there is a difficulty related to the possible singulathe contained
%in the  vertex function $\hat Q$.

   For exotic nuclei, we may need to employ a model space
consisting of two major shells. An example is the oxygen isotopes
with large neutron excess. In this case we may need to use
a two-major-shell model space consisting of the $0d1s0f1p$ shells.
There is a subtle singularity difficulty associated with such
multi-shell effective interactions.

\begin{figure}
\begin{center}
\scalebox{0.64}{\includegraphics[angle=0]{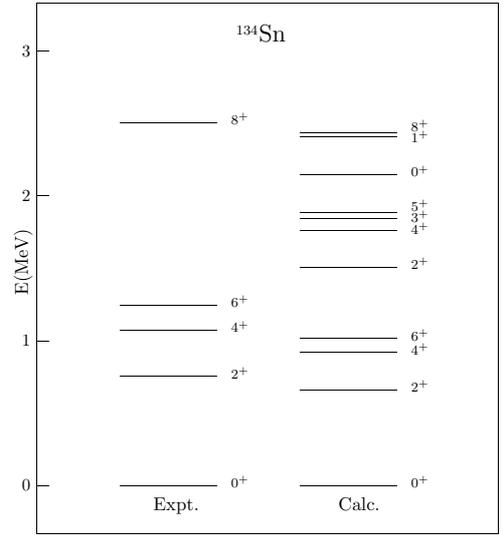}}
\caption{  Low-lying states of $^{134}Sn$
calculated with the folded-diagram $V_{eff}$ using 
the $V_{lowk}$ intraction. See section III for further
explanations.}
\end{center}
\end{figure}

Consider the core-polarization diagram d7 of Fig.\ 4.
Let us label its initial state by $(a,b)$, final state by $(c,d)$
and intermediate state by $(c,p,h,b)$. Here    $p$ and $h$  
denote respectively
the particle and hole lines of the diagram.
The energy denominator for this diagram is
$\Delta= (\epsilon_a - \epsilon_c) 
        -(\epsilon_p - \epsilon_h)$, $\epsilon$ being the 
single-particle energy.
This $\Delta$ may be equal to zero, causing the $\hat Q$-box to be singular.
For example this happens for $(a,b)=(0f,0f),~(c,d)=(0d,0d),~
p=0d,~h=0p$ when   a harmonic oscillator $H_0$ is used.
This singularity is not convenient for  the calculation of $V_{eff}$.
  It is remarkable and interesting
that these potential divergences can be circumvented by the recently
proposed $Z$-box method \cite{dong14,okamoto10,suzuki11}.
In this method, a vertex function $\hat Z$-box is employed. It is
related to the $\hat Q$-box by
\begin{equation}
\hat Z(\omega)=\frac{1}{1-\hat Q_1(\omega)}[\hat Q(\omega)
-\hat Q_1 (\omega)P(\omega-H_0)P],
\end{equation}
where $\hat Q_1$ is the first-order derivative of the $\hat Q$-box.
The $\hat Z$-box considered by Okamoto {\it et al.} \cite{okamoto10} is for
degenerate model spaces ($PH_0P=W_0$), while we consider here a more general
case with non-degenerate  $PH_0P$.
An important property of the above $\hat Z$-box is that it is finite when
the $\hat Q$-box is singular (has poles). Note that $\hat Z(\omega)$ 
satisfies
\bq
\hat Z(\omega) \chi _m =\hat Q(\omega)\chi_m
  ~{\rm at}~ \omega=E_m(\omega).
\eq 
Thus the converged $V_{eff}$ given by the $\hat Z$-box
is the same as by the $\hat Q$-box. In principle, we may use
either for its calculation. But the $\hat Z$-box is more convenient for
the calculation, especially for the multi-shell non-degenerate case
 where the $\hat Z$-box is a well-behaved function while 
the $\hat Q$-box
may have singularities \cite{dong14}.

\begin{figure}
\begin{center}
\scalebox{0.64}{\includegraphics[angle=0]{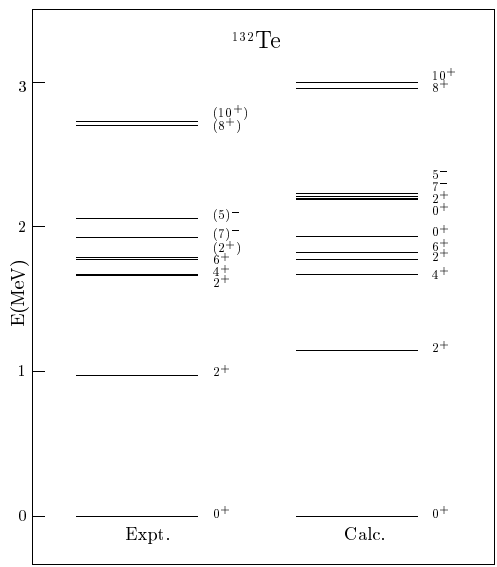}}
%\scalebox{0.75}{\includegraphics[angle=0]{corag136Te.eps}}
\caption{Same as Fig.\ 6 but for $^{132}Te$.}
\end{center}
\end{figure}

Let us now give a partial summary of what we have done.
 The formulations presented so far may seem  complicated. 
But the main point
is fairly simple. Namely we have shown that a general many-body problem 
specified by Hamiltonian $H$ can be formally reduced to a model-space problem 
consisting of a small number of quasi-particles confined
in a small model space with effective Hamiltonian $PH_{eff}P$. 
This Hamiltonian is given
by ($H_0$ + $V_{eff}$) where   $V_{eff}$ 
is the quasi-particle effective interaction which can be 
calculated from  the above
 folded-diagram series. Iterative methods for summing up 
this series for both degenerate and non-degenerate model spaces have been
formulated, as described earlier. 
This $V_{eff}$ method has been  successfully applied to shell-model 
nuclear structure
calculations \cite{jensen95,corag09}.

\begin{figure}
\begin{center}
%\scalebox{0.64}{\includegraphics[angle=0]{corag132Te.eps}}
\scalebox{0.75}{\includegraphics[angle=0]{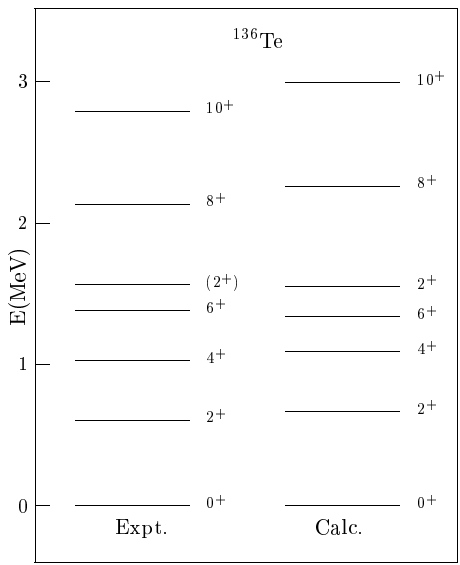}}
\caption{  Same as Fig.6 but for  $^{136}Te$.}
\end{center}
\end{figure}

  In the following section we  shall discuss the 
low-momentum NN interaction $V_{low-k}$
\cite{bogner01,bogner02,kuorg02,coraggio02,coragsn02,
corag03,bognerks03,schwenk02,holtct04,holtfam04}
which will be employed in evaluating  the above
folded-diagram $V_{eff}$ for shell-model calculations. This approach
has indeed been rather successful. As a preview, we display
some sample  results \cite{corag09} in Figs.\ 6-8.

\section{Low-momentum NN interaction }

In microscopic nuclear structure calculations starting from
a fundamental nucleon-nucleon (NN) potential $V_{NN}$, a well-known ambiguity
has been  the choice of the NN potential. There are a number
of successful models for $V_{NN}$, such as the CD-Bonn \cite{cdbonn},
Argonne V18 \cite{argonne}, Nijmegen \cite{nijmegen}
and Idaho \cite{chiralvnn} potentials. A common feature of these
potentials is that they all  
reproduce  the empirical deuteron properties and low-energy
phase shifts very accurately. But, as illustrated in Figs.\ 1-2
which show the momentum-space matrix elements, 
 these potentials are in fact quite different from each other.
This has been a situation of much concern: certainly we would like
to have a ``unique" NN potential.  If there does exist 
one unique NN potential,
then one is confronted with the difficult question
in deciding which, if any,  of the existing potential models is the
correct one. In the past several years, there have been extensive studies
of the so-called low-momentum NN potential $V_{low-k}$ 
\cite{bogner01,bogner02,kuorg02,coraggio02,coragsn02,
corag03,bognerks03,schwenk02,holtct04,holtfam04}.
%\cite{bogner01,bogner02,kuorg02,coraggio02,coragsn02,corag03,bognerks03,
%schwenk03,holtct05,holtfam04}.
In the following, we shall discuss this development in some detail.

The potential $V_{low-k}$ is based on  the renormalization group (RG) and 
effective field theory (EFT) approaches 
\cite{weinberg,kaplan,kaplan98,lepage,epel98,haxton}.
 A central theme of the RG-EFT approaches 
is that in probing physics in the infrared region, it is generally adequate 
to employ a low-energy (or low-momentum) effective 
theory. One starts with a chosen underlying theory, in the present case one
of the high-precision nucleon-nucleon potential models.
There are states of both high and low energy in the theory. 
Then by a decimation procedure where the high-energy modes
are integrated out, one obtains an
 effective field theory for the low-energy modes. How to carry out this integration
is of course an important step. Here the methods of the renormalization
group  come in. For low-energy physics, one usually employs
a model space below
a cut-off momentum $\Lambda$, which is typically several hundred MeV.
All fields with momentum greater than $\Lambda$ are  
integrated (or decimated) out,
and in this way we obtain an EFT for momentum within $\Lambda$.

It is tempting to apply the above RG-EFT idea to realistic NN potentials.
These models all have a built-in strong short-range repulsion, 
as is well known.  As a result, they all have strong high-momentum components.  
We are quite 
certain about the physics of the long-range (or low-momentum) part of $V_{NN}$.
After all, at the largest distances the nucleon-nucleon potential is essentially just 
one-pion exchange. However, we are less certain
about the very short-range (or high-momentum) part; in most models it is
put in by hand 
phenomenologically. Thus the high-momentum modes of $V_{NN}$ are
model dependent. In line with the above RG-EFT approach, 
it seems to be both of interest and desirable to just ``integrate out" 
the short-range or the high-momentum part of the NN potential, thereby 
obtaining a low-momentum effective potential $V_{low-k}$. In so doing we
are indeed trying to weed out the most uncertain parts of the NN potential.
In the following, let us first describe our method for carrying out the above
integration. The $V_{low-k}$ derived from different NN potentials
will be compared. We shall show that for a specific cut-off momentum
$\Lambda$ the $V_{low-k}$ derived from  different NN potential models
are nearly identical to each other. In addition, it is a smooth potential
and appears to be suitable for directly computing shell-model effective
interactions.
% without first calculating the Brueckner G-matrix.

\vskip 0.3cm
{\bf III.A. $T$-matrix equivalence approach}
\vskip 0.3cm
 
 We now discuss how to calculate $V_{low-k}$. In integrating out
the  high-momentum parts mentioned earlier, an important
requirement is that certain low-energy physics of $V_{NN}$
is exactly preserved by $V_{low-k}$. For the two-nucleon problem,
there is one bound state, namely the deuteron. Thus one must require
the deuteron properties given by $V_{NN}$ to be preserved by $V_{low-k}$.
In nuclear effective-interaction theory, there are several well developed
model-space reduction methods. One of them is the Kuo-Lee-Ratcliff (KLR)
folded-diagram method \cite{kuos,klr71}, which was originally formulated
for discrete (bound state) problems. We have discussed this method
in detail earlier in section II.

For the two-nucleon problem, we want the effective interaction $V_{low-k}$
to preserve  also the low-energy scattering phase shifts, in addition  to
the deuteron binding energy. Thus we need an effective interaction
for scattering (unbound) problems. A convenient framework 
for this purpose
is a $T$-matrix equivalence approach as described below.

We use a  continuum model space specified by $k \leq \Lambda $, namely
\begin{equation}
P=\int _0 ^{\Lambda} d\vec k \; | \vec k  \rangle \langle \vec k |,
\end{equation} 
where $k$ is the two-nucleon relative momentum and $\Lambda$ is the cut-off
momentum which is also known as the decimation momentum. Its typical value 
is about 2 $fm^{-1}$ as we shall discuss later. Our purpose is to look
for an effective interaction $PV_{eff}P$, with $P$ defined above, which
preserves certain properties of the full-space interaction $V_{NN}$ for both
bound and unbound states. This effective interaction will be referred to
as $V_{low-k}$.

 We start from the full-space half-on-shell $T$-matrix
\br
&&  T(k',k,\omega)  = V_{NN}(k',k) \\   
&&  + {\cal P} \int _{0} ^{\infty} q^2 dq  V_{NN}(k',q) 
 \frac{1}{\omega -H_0(q) } T(q,k,\omega) , \nonumber
\er
where $\omega=\varepsilon _k$.
This is the $T$-matrix for the two-nucleon problem with Hamiltonian
$H=H_0+V_{NN}$ and $H_0$ represents the relative kinetic energy
with its eigenvalue denoted by $\varepsilon _k$. The symbol ${\cal P}$  
 denotes the principle value integration.

We then define a $P$-space low-momentum $T$-matrix by 
\begin{eqnarray}
 && T_{low-k }(p',p,\omega)  =   V_{low-k}(p',p)  \\
  &&\hspace{-0.6in}+ {\cal P}\int _{0} ^{\Lambda} q^2 dq  V_{low-k}(p',q) 
 \frac{1}{\omega -H_0(q) } T_{low-k} (q,p,\omega) , \nonumber 
\end{eqnarray}
where  $\omega=\varepsilon _p$, $(p',p) \leq \Lambda$,
 and the integration  is from 0 to $\Lambda$.
We require the equivalence condition
\begin{equation}
 T(p',p,\varepsilon_ p) 
= T_{low-k}(p',p, \varepsilon_ p) ;
~p',p\leq \Lambda.
\end{equation}
The above equations define  the effective low-momentum interaction; it is 
required to preserve the low-momentum ($\leq \Lambda$) 
half-on-shell $T$-matrix. Since phase shifts are given by the fully-on-shell
$T$-matrix $T(p,p,\varepsilon_ p)$, low-energy phase shifts given by
the above $V_{low-k}$ are clearly the same as those of $V_{NN}$.

 In the following, let us show that a solution of the above equations
may be found by way of a folded-diagram  method 
\cite{bogner01,klr71,kuos}.  
 The $T$-matrix of Eq.\ (37) can be written as
\begin{equation}
 \langle k' 
| (V +V\frac{1}{e(\varepsilon _k)}V 
+V\frac{1}{e(\varepsilon _k)}V\frac{1}{e( 
\varepsilon _k)}V+
\cdots) | k \rangle , \nonumber 
\end{equation}
 where $e(\omega )\equiv (\omega - H_o)$. (For simplicity, we have used
$V$ to denote $V_{NN}$.)
 Note that the intermediate states (represented by 1 in the numerator)
cover the entire space. In other words, we have $1=P+Q$ where $P$ 
denotes the model space (momentum $\leq \Lambda$) and $Q$ its complement. 
Expanding it out in terms of $P$ and $Q$ and defining
a  $\hat Q$-box as 
\begin{eqnarray}
&& \hspace{-.2in} \langle k' | \hat Q(\omega) | k \rangle \\
&&=  \langle k' | [V
 +V\frac{Q}{e(\omega )}V+V\frac{Q}{e(\omega)}
V\frac{Q}{e(\omega ) }V+ \cdot \cdot \cdot] | k \rangle, \nonumber
\end{eqnarray}
one readily sees that the $P$-space portion of 
the $T$-matrix can be regrouped as a $\hat Q$-box
series, namely 
\begin{eqnarray}
&& \hspace{-.4in} \langle p' | T(\omega ) | p \rangle = 
\langle p' | [
\hat Q(\omega)  +\hat Q(\omega) \frac{P}{e(\omega)} \hat Q(\omega) \\
& & + \hat Q(\omega) \frac{P}{e(\omega)} \hat Q(\omega)  
\frac{P}{e(\omega)} \hat Q(\omega)
 +\cdot \cdot \cdot
] | p \rangle , \nonumber
\end{eqnarray} 
where $\omega=\varepsilon_p$. Note that the intermediate states of each $\hat Q$-box all belong to the 
Q-space. Denoting each $\hat Q$ by a circle, the  above
$T$-matrix is given by the sum of all the terms in the left
column of Fig.\ 9 (namely $T=A+B+C+\cdots$).
\begin{figure*}[t]
\begin{center}
%\scalebox{0.45}{\includegraphics{luigifold.eps}}
\scalebox{0.45}{\includegraphics{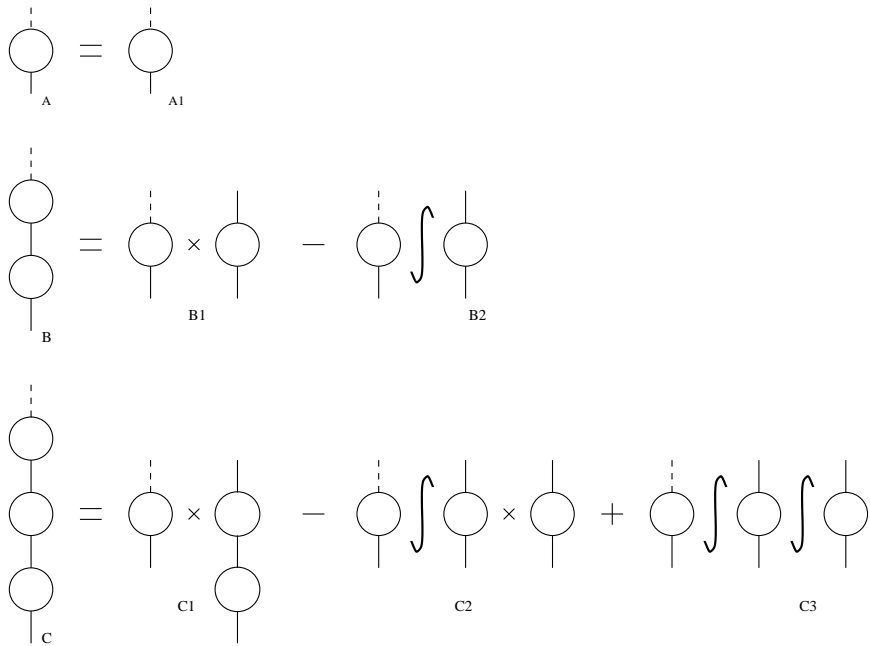}}
\caption{Expansion of $T$-matrix in terms of $\hat Q$-box (represented
by a circle).
Each integral sign represents a fold}\label{figure1}
\end{center}
\end{figure*}

All the $\hat Q$-boxes in the above equation
have the same energy variable, namely $\varepsilon _p$. Let us introduce
a folded-diagram factorization \cite{kuos}: We factorize the 
two-$\hat Q$-box term as 
\begin{eqnarray}
&& \hspace{-.3in}\langle p' | \hat Q(\varepsilon _p ) \frac{P}{e(\varepsilon _p )}
 \hat Q(\varepsilon _p) | p \rangle = \\
&& \hspace{-.1in} \sum _{p''} \langle p' | \hat Q(\varepsilon _p'' )
| p'' \rangle \langle p'' |
 \frac{P}{e(\varepsilon _p )} \hat Q(\varepsilon _p) | p \rangle
 -\langle p' | \hat Q'\int \hat Q | p \rangle \nonumber ,
\end{eqnarray} 
where the last term is the once-folded two-$\hat Q$-box term.
The folded term is simply the difference between the original and
factorized two-$\hat Q$-box terms.
Schematically the above factorization is written as
\begin{equation}
\hat Q \frac{P}{e} \hat Q=
\hat Q\times   \hat Q- \hat Q' \int \hat Q.
\end{equation}
Diagrammatically, this equation is represented by the second equation
of Fig.\ 9, where (B) represents the lhs two-$\hat Q$-box term, (B1)
the factorized two-$\hat Q$-box term and (B2) the corresponding folded
term. A subtle difference between the vertex functions $\hat Q$ and
$\hat Q'$ may be pointed out. 
 $\hat Q$ has diagrams with one, two, three,... $V$ vertices. Those with
one $V$ vertex are energy independent, and for them there is 
no folded-diagram correction. 
Thus $\hat Q'$ is the same as $\hat Q$ except with
its diagrams first order in $V$ removed. This relation remains the same
for higher order folded diagrams.

Similarly we can factorize the three-$\hat Q$-box term as 
\br
\hat Q \frac{P}{e} \hat Q \frac{P}{e} \hat Q  
&=&\hat Q\times   \hat Q\frac{P}{e}\hat Q 
- [\hat Q' \int \hat Q] \times \hat Q \nonumber \\
&& + \hat Q' \int \hat Q \int  \hat Q,
\er
where the last term is a twice-folded three-$\hat Q$-box term. 
Diagrammatically this equation is represented by the third equation
of Fig.\ 9, where (C) denotes the lhs three-$\hat Q$-box term while
the corresponding factorized, factorized and once-folded, and twice-folded
terms are represented respectively by diagrams (C1), (C2) and (C3). 

The folded-diagram factorization for the four-$\hat Q$-box term of $T$ can 
be carried out in a similar way.
And this four $\hat Q$-box term of $T$ is factorized into four terms
(D1) to (D4). (These terms are not shown in Fig.\ 9, but their structure
is closely similar to  the three-$\hat Q$-box factorization.
For example, (D4) is of the form
-$\hat Q\int \hat Q\int \hat Q\int \hat Q$.)
 Continuing this procedure, a simple
structure becomes transparent. The sum of the diagrams 
(A), (B2), (C3), (D4),... of Fig.\ 9 is just the effective interaction
\begin{eqnarray}
V_{low-k}& =& \hat{Q} - \hat{Q'} \int \hat{Q} + \hat{Q'} \int \hat{Q} \int 
\hat{Q} \nonumber \\
&&- \hat{Q'} \int \hat{Q} \int \hat{Q} \int \hat{Q} + \cdots
\end{eqnarray}
The sum of the diagrams (B1), (C2), (D3), ... is just 
$V_{low-k} \frac{1}{e}\hat Q$. Since $T$ is the sum of the diagrams 
(A), (B), (C), (D), ... it is readily seen that by regrouping the diagrams 
of Fig.\ 9 we have the $T$-matrix equation
 $T=V_{low-k}+V_{low-k}\frac{1}{e}T$. Namely the  $V_{low-k}$ given by the
above equation is a solution of Eqs.\ (30-32).

Note that this is just the KLR folded-diagram effective interaction 
\cite{kuos,klr71} as given earlier in Eq.\ (21). The KLR method 
was originally formulated for
bound-state problems. We now see that  it is also applicable to
scattering problems; it preserves the half-on-shell $T$-matrix.
This implies the preservation of  not only the low-energy phase shifts 
(which are given by the
fully-on-shell $T$-matrix) but also the low-momentum components of the
scattering wave functions.

We have shown above that certain low-energy physical quantities given by
the full-momentum potential $V_{NN}$ are preserved by the low-momentum
potential $V_{low-k}$. This preservation is an important point, 
and it should be 
numerically checked, to see for instance if the deuteron binding energy
and low-energy phase shifts given by $V_{NN}$ are indeed reproduced
by $V_{low-k}$.
We have checked the deuteron binding energy $BE_d$ given by $V_{low-k}$.
For a range of $\Lambda$, such as $0.5 fm^{-1}\leq \Lambda \leq 3 fm ^{-1}$,
$BE_d$ given by $V_{low-k}$ agrees very accurately (to 4 places after the
decimal point) with that given by $V_{NN}$. 
We have also checked
the phase shifts and the $T$-matrix $T(p',p,\omega = p^2)$ with $(p',p)\le
\Lambda$; very good agreement between these quantities given by
$V_{NN}$ and $V_{low-k}$ were obtained.

An important question
in our approach is  the choice of the momentum cut-off $\Lambda$.
Phase shifts are given by the 
fully-on-shell $T$-matrix, $T(p,p, \varepsilon _p)$. 
Hence for a chosen $\Lambda$, $V_{low-k}$ can only produce
phase shifts  up to $E_{lab}=2 \hbar ^2 \Lambda^ 2/M$, $M$ being the nucleon
mass.  Realistic NN potentials are constructed to fit empirical phase
shifts up to $E_{lab}\approx 350$ MeV \cite{cdbonn}. 
 It is reasonable then to require our $V_{low-k}$ to reproduce phase shifts
also up to this energy. Thus one should use $\Lambda$ in the vicinity
of 2 $fm^{-1}$.
\begin{figure}
\begin{center}
%\scalebox{0.33}{\includegraphics[angle=-90]{715vlowk1s01s0.eps5v}}
\scalebox{0.33}{\includegraphics[angle=-90]{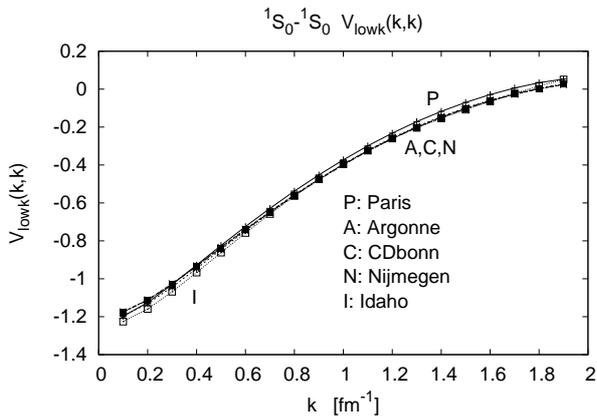}}
%figureeps
\caption{  Comparison of $^1S_0-{^1S_0}$  
 diagonal momentum-space matrix elements
$V_{low-k}(k,k)$ of Paris \cite{paris}, 
CD-Bonn \cite{cdbonn}, 
Argonne \cite{argonne}, 
Nijmegen \cite{nijmegen} 
and Idaho \cite{chiralvnn} potentials.}
\end{center}
\end{figure}

In Figs.\ 10-11, we compare some $V_{low-k}(k,k)$ matrix elements
calculated from several $V_{NN}$ potentials. 
A decimation momentum of $\Lambda=2.0fm^{-1}$ is used in
the $V_{low-k}$ calculation.
It is seen that
the matrix elements given by the various potentials are nearly
identical, in sharp contrast to Figs.\ 1-2 where the 
corresponding $V_{NN}(k,k)$
matrix elements  are vastly different. This is an interesting
result, suggesting that
 the $V_{low-k}$ interactions derived for different 
realistic $V_{NN}$ potentials are nearly uinique 
\cite{bognerks03}. Note that
these $V_{NN}$ potentials all reproduce the low-energy
two-nucleon experimental data (deuteron binding energy
and NN phase shifts up to $E_{lab}\simeq 350$ MeV) very well.

\begin{figure}
\begin{center}
%\scalebox{0.33}{\includegraphics[angle=-90]{715vlowk3s13s1.eps5v}}
\scalebox{0.33}{\includegraphics[angle=-90]{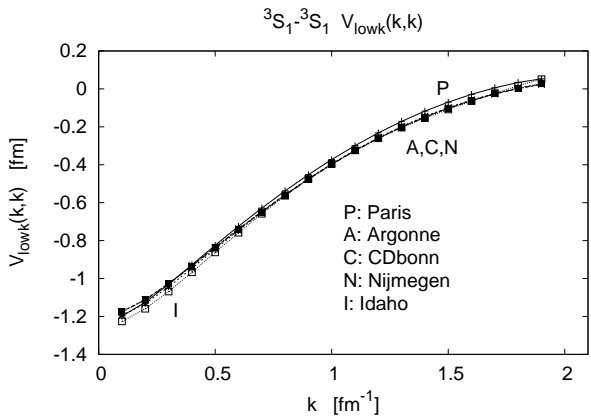}}
%figureeps
\caption{  Same as Fig.\ 10 but for the
 $^3S_1-{^3S_1}$ channel.
}
\end{center}
\end{figure}

Extensive shell-model calculations
based on $V_{low-k}$ have been carried out with rather
encouraging results (see e.g., Ref.\ \cite{corag09}). To
illustrate, we compare in Figs.\ 7-9 the calculated 
spectra with the
experimental ones for three nuclei
 $^{134}Sn$, $^{132}Te$ and $^{136} Te$. The $V_{low-k}$
obtained from the CD-Bonn potential \cite{cdbonn}
using a decimation momentum $\Lambda=2.0 fm^{-1}$
has been employed \cite{corag09}. As seen the calculated spectra
agree with the experimental ones rather satisfactorily.

For many years, the Brueckner $G$-matrix interaction was employed
in nuclear calculations using realistic NN interactions \cite{jensen95}.
Both $G$-matrix and $V_{low-k}$ transform, or tame, the
NN interactions with strong repulsive cores 
into  smooth interactions without such cores, the latter being
much simpler for calculations than the former. 
A difference between these two interactions may be mentioned.
The $G$-matrix  
interaction is energy dependent while the $V_{low-k}$ is not.
This makes the latter more convenient in applications 
\cite{corag09,holt05,bogner10,hebeler15}.

\vskip 0.3cm
{\bf III.B. Counter terms}
\vskip 0.3cm

A main point of the RG-EFT theory is that  low-energy physics
is not sensitive to  fields beyond a cut-off scale $\Lambda$.
Thus for treating low-energy physics, one just integrates out
the fields beyond a cut-off scale $\Lambda$, thereby 
obtaining a low-energy
effective field theory. In RG-EFT, this integrating out, or decimation,
 generates an infinite series of counter terms \cite{lepage} 
which are  simple power series in momentum. This is a very useful
and interesting result. When we derive our
low-momentum interaction, the high-momentum modes of the input interaction
are integrated out. Does this decimation also generate a series of counter 
terms?  If so, what are the counter terms so generated? Holt, Kuo, Brown
and Bogner \cite{holtct04} have studied this question, 
and  we would
like to review their results here. 

\begin{figure}
\begin{center}
\scalebox{0.42}{\includegraphics[angle=0]{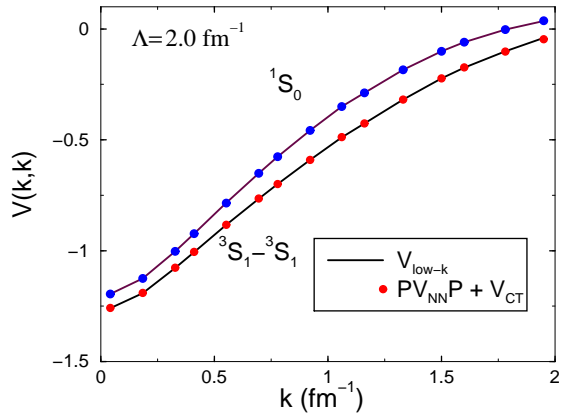}}
%\centerline{\epsfxsize=3.0in\epsfbox{ctholtfig1.eps}}   
\caption{Comparison of 
$V_{low-k}$ with $PV_{NN}P$ plus counter terms
for $^1S_0$ and $^3S_1$ channels. \label{inter}}
\end{center}
\end{figure}

Similar to the usual counter term approach, we assume that the difference
between $V_{low-k}$ and $V_{NN}$ can be accounted for by a series
of counter terms. Specifically, we consider
\br
 V_{low-k}(q,q')&\simeq& V_{NN}(q,q') \nonumber \\
&& +V_{counter}(q,q'); 
~(q,q')\leq \Lambda,
\er
where $V_{NN}$ is the free-space NN potential from which $V_{low-k}$
is derived  and
the counter term potential is given as a power series 
\begin{eqnarray}
&&\hspace{-.3in}V_{counter}(q,q')  = \nonumber \\
&& C_0+C_2q^2+C_2'q'^2+C_4(q^4+q'^4)+C_4'q^2q'^2  \nonumber \\
                  &   & +C_6(q^6+q'^6)+C_6'q^4q'^2+C''_6q^2q'^4+....
\end{eqnarray}

\begin{figure}
\begin{center}
\scalebox{0.35}{\includegraphics[angle=-90]{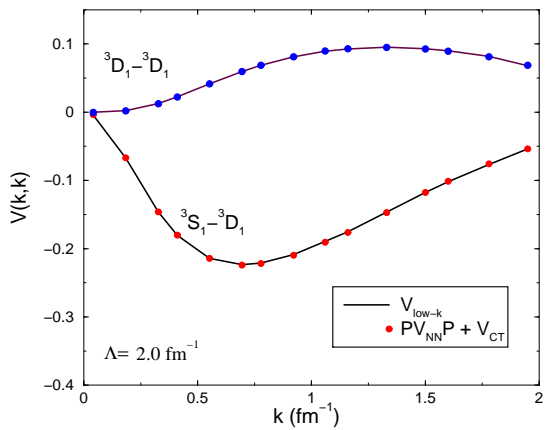}}
\caption{Same as Fig. 12  but 
for $^3D_1$ and $^3S_1-^3D_1$  channels. \label{inter}}
\end{center}
\end{figure}

The counter term coefficients are determined using standard fitting 
techniques 
so that the right hand side of Eq.~(48) provides a best fit to the left
hand side of the same equation. We perform this fitting over all partial
wave channels, and find consistently good agreement. In Fig.~12 
we compare some $^1S_0$ and $^3S_1$ matrix elements of 
$(PV_{NN}P+V_{CT})$ 
with those of $V_{low-k}$ for momenta below the cutoff $\Lambda$. 
 Here $P$ denotes the projection
operator for states with momentum less than $\Lambda$.
A similar comparison for $^3D_1$ and $^3S_1-{^3D_1}$ channels
is displayed in Fig.~13.
The agreement is indeed very good, and also for phase shifts 
as illustrated in Fig.~14.

Now let us examine the counter terms themselves.  In Table 2, we list some 
of the counter term coefficients, using CD-Bonn as our bare potential. 
In the table we list only the counter terms for the $^1S_0$ and 
$^3S_1-{^3D_1}$ 
partial waves; we have found that the counter terms for all the other 
waves 
are much smaller. This tells us an interesting result, namely, except for 
the 
above two channels, $V_{low-k}$ is very similar to $V_{NN}$ alone. We 
also point out that the coefficients $C_6$ are found to be very small, 
indicating that the above power series expansion converges rapidly. 
 In the last
row of the table, we list the rms deviations between $V_{low-k}$ and 
$PV_{NN}P+V_{counter}$; the fit is indeed very good.

% We note that the low-momentum NN potential given by Eqs.(1-4)
% is not Hermitian. Our numerical results are obtained using
%a Hermitian version of   $V_{low-k}$ calculated with  
%the  Okubo transformation method as to be discussed in Section 3.
%The counter terms obtained for the interaction of Eq.(4) and 
%those for the Hermitian one
%are  in fact quite similar to each other.

As shown in the table, the counter terms are all rather small except
for $C_0$ and $C_2$ of the $S$ waves. This is consistent with 
the RG-EFT approach where the
counter term potential is given as a delta function plus its derivatives
\cite{lepage}. 

\begin{figure}
\begin{center}
\scalebox{0.35}{\includegraphics[angle=-90]{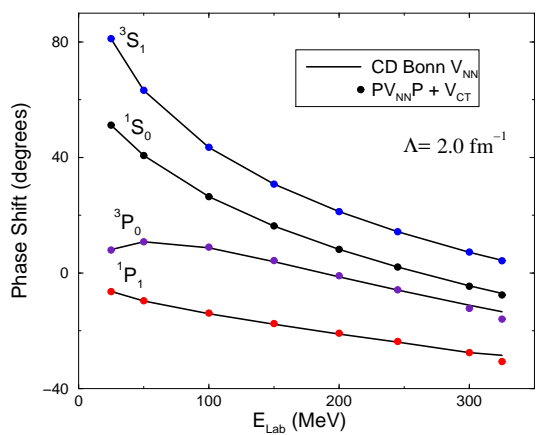}}
\caption{Comparison of phase shifts calculated from 
$V_{low-k}$ with those from $PV_{NN}P$ plus counter terms.}
\end{center}
\end{figure}

Comparing counter term coefficients for different $V_{bare}$ potentials 
can 
illustrate key differences between those potentials. For example, we have found
that the $^1S_0$ $C_0$ coefficients for the CD-Bonn \cite{cdbonn}, 
Nijmegen \cite{nijmegen}, Argonne \cite{argonne} and Paris \cite{paris} NN 
potentials are respectively -0.158, -0.570, -0.753
and -1.162. Similarly, the $^3S_1$ $C_0$ coefficients for these
potentials are respectively -0.467, -1.082, -1.148 and -2.224.
That the $C_0$ coefficients for these potentials are significantly 
different is a reflection that the short-range repulsions built into these 
potentials are different. For instance, the Paris potential effectively 
has a 
very strong short-range repulsion and consequently its $C_0$ is much 
larger than the $C_0$ of the others. The short-range repulsion contained
in $NN$
potentials is uncertain and model dependent. Further study
is needed for its understanding. 

\begin{table}
\caption{Coefficients of the counter terms for $V_{low-k}$
obtained from the  CD-Bonn potential using 
$\Lambda = 2fm^{-1}$. The unit for the combined quantity $C_n k^n$
is $fm$, with momentum $k$ in units of $fm^{-1}$. }
%\vskip 0.5cm 
{\begin{tabular}{@{}cccccc@{}}
%{|cp{1.0in}|cp{1.0in}|cp{1.0in}|cp{1.0in}|cp{1.0in}}
\hline
          & $^1S_0$  & $^3S_1$ & $^3S_1-^3D_1$ & $^3D_1$ &       \\
\hline
$ C_0  $  & -0.1580   & -0.4646  &     0    &     0    &  \\ 
$ C_2  $  & -0.0131   &  0.0581  & -0.0017  & -0.0005  &  \\ 
$ C_2' $  & -0.0131   &  0.0581  &  0.0301  & -0.0005  &  \\ 
$ C_4  $  &  0.0004   & -0.0011  & -0.0013  &  0.0006  &  \\ 
$ C_4' $  & -0.0011   & -0.0113  & -0.0047  & -0.0018  &  \\ 
$ C_6  $  & -0.0004   & -0.0004  &  0.0006  & -0.0001  &  \\ 
$ C_6' $  & -0.0005   &  0.0005  & -0.0001  & -0.0001  &  \\ 
$ C_6''$  & -0.0005   &  0.0005  & -0.0003  & -0.0001  &  \\ 
\hline
$\Delta _{rms}$ & 0.0002 & 0.0003  & 0.0028 &  0.0003  & \\       
\hline
\end{tabular}}
\end{table}

 As shown in the Table, the $C$ coefficients for the \hskip 1cm
$^3S_1-{^3D_1}$
channel are all very small. Thus the counter term for 
the NN tensor force is nearly
vanishing   and consequently  
$V_{low-k}(tensor) \simeq V_{NN}(tensor)$.
It appears that
the tensor force is exempted from  renormalization. 
It should be of interest to further study this possible exemption
property of the tensor force.

%Tsunoda et al. \cite{tsunoda11}

\vskip 0.3cm
{\bf III.C.  Hermitian low-momentum interactions}
\vskip 0.3cm
 It should be pointed out that the low-momentum NN interaction given by the 
above folded-diagram expansion is not Hermitian. This is not 
a desirable feature,
 and one would like to have  an interaction which is Hermitian.
Methods for deriving Hermitian effective interactions have been developed by
Okubo \cite{okubo} and Andreozzi \cite{andre96}.
 A general framework for constructing Hermitian effective interactions
was recently studied by Holt, Kuo and Brown \cite{holtfam04}.
In this section, we shall discuss their method
for  derivation of  Hermitian effective interactions.
As we shall see shortly, one can in fact construct a family of Hermitian
low-momentum NN interactions that are phase-shift equivalent.

Let us denote the folded-diagram  low-momentum NN interaction 
as $V_{LS}$.  (Recall that we have used the Lee-Suzuki (LS) method for 
its derivation.)
$V_{LS}$ preserves the half-on-shell $T$-matrix $T(k',k,k^2)$ for
$(k',k) \leq \Lambda$. If we relax this half-on-shell constraint,
we can obtain low-momentum NN interactions which are Hermitian.
(Note the Hermitian potentials discussed below all preserve
the fully on-shell $T$-matrix and are phase-shift equivalent
 \cite{holtfam04}.)

  The model-space secular  equation for $V_{LS}$ is
\begin{equation}
P(H_0+V_{LS})P\chi _m =E_m \chi _m,
\end{equation}
where $\{E_m\}$ is a subset of the eigenvalues $\{E_n\}$ of the
full-space Schroedinger equation $(H_0+V)\Psi _n =E_n \Psi _n$.
$\chi _m$ is the $P$-space projection of the full-space wave function 
$\Psi _m$, namely
$ \chi _m =P \Psi _m $. 
The above effective interaction may be rewritten in terms
of a wave operator $\omega$, namely
\begin{equation}
PV_{LS}P =Pe^{-\omega}(H_0 +V)e^{\omega}P-PH_0P,
\end{equation}
where $\omega$ possesses the usual properties:
$\omega = Q\omega P; 
\chi_m = e^{-\omega} \Psi _m ; 
\omega \chi_m = Q\Psi _m.$ Here $Q$ is the complement of $P$, $P+Q=1$.

While the full-space eigenvectors $\Psi _n$  are orthogonal to each other,  
 the model-space eigenvectors $\chi _m$  are clearly not so and 
consequently the 
effective interaction $V_{LS}$ is not Hermitian. We now make a 
$Z$ transformation such that 
\begin{eqnarray}
Z \chi_m &=& v _m; \nonumber \\
\langle v_m \mid v_{m'} \rangle &=& \delta _{mm'};~m,m'=1,d,
\end{eqnarray}
where $d$ is the dimension of the model space. This transformation
reorients the vectors $\chi_m$ such that they become orthonormal
to each other.
We assume that $\chi _m$'s ($m=1,...,d$)  are linearly independent so that
$Z^{-1}$ exists, otherwise the above transformation is not possible.  
Since $v_m$ and $Z$ exist entirely within the model space, we can write
$v_m=Pv_m$ and $Z=PZP$. 

Using Eq.\ (52), we transform Eq.\ (50) into
\begin{equation}
Z(H_0+V_{LS})Z^{-1}v_m=E_m v_m,
\end{equation}
which implies
\begin{equation}
Z(H_0+V_{LS})Z^{-1}=\sum _{m\epsilon P} E_m | v_m \rangle 
\langle v_m |.
\end{equation}
Since $E_m$ is real (it is an eigenvalue of $(H_0+V)$ which is Hermitian) 
and the vectors $v_m$
are orthonormal to each other, 
$Z(H_0+V_{LS})Z^{-1}$ is clearly Hermitian.  
The non-Hermitian secular equation, Eq.\ (12), is now transformed into a 
Hermitian model-space eigenvalue problem
\begin{equation}
P(H_0 +V_{herm})Pv_m=E_m v_m
\end{equation}
with the Hermitian  effective interaction 
\begin{equation}
V_{herm}=Z(H_0+V_{LS})Z^{-1}-PH_0P,
\end{equation}
or equivalently
\begin{equation}
V_{herm}=Ze^{-\omega}(H_0+V)e^{\omega}Z^{-1}-PH_0P.
\end{equation}

To calculate $V_{herm}$, we must first have the $Z$ transformation. 
Since there are certainly many ways to construct $Z$, this generates
a family of Hermitian effective interactions, all originating from $V_{LS}$.
For example, we can construct $Z$ using the familiar Schmidt 
orthogonalization procedure, namely:
\begin{eqnarray}
v_1&=&Z_{11}\chi _1  \nonumber  \\
v_2&=&Z_{21}\chi _1 +Z_{22} \chi _2  \nonumber  \\
v_3&=&Z_{31}\chi _1 +Z_{32} \chi _2  +Z_{33} \chi _3 \nonumber  \\
v_4&=&......,
\end{eqnarray}
with the matrix elements $Z_{ij}$ determined from Eq.~(52). 
We denote the Hermitian effective interaction using this $Z$ 
transformation as $V_{schm}$.  Clearly there is 
more than one such Schmidt procedure. For instance, we can 
use $v_2$ as the starting point, which gives
$v_{2} = Z_{22}\chi _2$, $v_3=Z_{31} \chi _1+Z_{32}\chi_2$,
and so forth. This freedom in choosing the orthogonalization procedure
actually gives us  many ways to generate a Hermitian 
interaction, and this is our family of Hermitian interactions produced from
$V_{LS}$.

We now show how some well-known Hermitization transformations relate to 
(and in fact, are special cases of) ours.
We first look at the Okubo transformation \cite{okubo}. From the properties
of the wave operator $\omega$,  we have
\begin{equation}
\langle \chi _m | (1+\omega ^+ \omega ) | \chi _{m'} \rangle
=\delta _{mm'}.
\end{equation}
It follows that an analytic choice for the $Z$ transformation is
\begin{equation}
Z=P (1+\omega ^+ \omega)^{1/2}P.
\end{equation}
This leads to the Hermitian effective interaction
\br
&&\hspace{-.3in}V_{okb-1}= P (1+\omega ^+ \omega)^{1/2}P(H_0+V_{LS})P
 (1+\omega ^+ \omega)^{-1/2}P \nonumber \\
&& \hskip 0.8cm -PH_0P.
\er
 It is easily seen that the above is
equal to the Okubo Hermitian effective interaction
\br
&&\hspace{-.4in}V_{okb}=P (1+\omega ^+ \omega)^{-1/2}(1+\omega^+)(H_0+V)
  \nonumber \\
& & \times (1+\omega) (1+\omega ^+ \omega)^{-1/2}P-PH_0P,
\er
giving us an alternative expression, Eq.~(62), for the Okubo interaction.

There is another interesting choice for the transformation $Z$. As pointed 
out by Andreozzi \cite{andre96}, the positive definite operator
$P(1+\omega ^+ \omega)P$ can be decomposed into two Cholesky matrices,
namely
\begin{equation}
P (1+\omega ^+ \omega)P= PL L^TP,
\end{equation}
where $L$ is a lower triangle Cholesky matrix, $L^T$ being its transpose.
Since $L$ is real and it is within the $P$-space, we have  
\begin{equation} 
Z=L^T 
\end{equation}
and the corresponding Hermitian effective interaction from Eq.\ (56) is
\begin{equation}
V_{cho}=PL^TP(H_0+V_{LS})P(L^{-1})^TP -PH_0P.
\end{equation}
This is the Hermitian effective interaction of Andreozzi \cite{andre96}.

The  Hermitian effective interaction  of Suzuki and Okamoto 
\cite{suzuki94,suzuki83} is of the form
\begin{equation}
V_{suzu}=Pe^{-G}(H_0+V)e^{G}P-PH_0P
\end{equation}
with $G=tanh^{-1}(\omega-\omega^{\dagger})$ and $G^{\dagger}=-G$.
It has been shown that this interaction is the same as the Okubo interaction 
\cite{suzuki83}.
In terms of the $Z$ transformation, it is readily seen that 
the operator $e^{-G}$  is equal to $Ze^{-\omega}$ 
with $Z$ given by Eq.\ (52). Thus, these three well-known 
Hermitian effective interactions indeed belong to our family.
The family of Hermitian potentials are all phase-shift
equivalent to $V_{LS}$ \cite{holtfam04}.

Using a solvable matrix model, one finds that 
the above Hermitian effective interactions
$V_{schm},V_{okb}$ and $V_{cho}$ can   be quite different
\cite{holtfam04} from each
other and from $V_{LS}$, especially when $V_{LS}$ is largely non-Hermitian.
For the $V_{NN}$ case, it is fortunate that the $V_{low-k}$ coresponding
to $V_{LS}$ is only slightly non-Hermitian.
As a result, the Hermitian low-momentum NN inteactions corresponding to 
$V_{schm},V_{okb}$ and $V_{cho}$ are all quite similar to each other
and to $V_{LS}$. Thus in applications it is the $V_{low-k}$ 
corresponding to $V_{LS}$ which is commonly used \cite{corag09,holtfam04}.

%\begin{figure}[ht]
%%\centerline{\epsfxsize=3.0in\epsfbox{hermfig1.eps}}   
%\caption{Comparison of 
%non-Hermitian (Lee-Suzuki) and
%Hermitian (Okubo, Cholesky(Andreozzi), Schmidt) low-momentum
%NN interactions. \label{inter}}
%\end{figure}

\section{Brown-Rho scaling and three-nucleon force}

As indicated by Figs.\ 6-8, the $V_{low-k}$ interactions
derived from free-space $V_{NN}$ have worked well 
in shell-model calculations involving
mainly valence nucleons. But for infinite symmetric nuclear matter, 
such free-space two-nucleon 
interactions alone are unable to reproduce simultaneously the 
empirical saturation energy and density ($E_0/A \simeq -16$\,MeV and 
$n_0 = 0.16$\,fm$^{-3}$)
\cite{siu09,dong09,bogner05,dongbrs13}.
To illustrate, we display in Fig.\ 15 results from
a recent nuclear matter calculation \cite{dongbrs13}
using a non-pertubative ring-diagram resummation 
method that will be outlined later.
There the lowest curve (C) is obtained with
the free-space BonnS potential \cite{bonns}. As seen this curve
descends rapidly with density, showing no sign of reaching a local
minimum at the saturation energy and density. 
In the present section we outline how this over-binding problem 
may be overcome by using density-dependent effective interactions
generated by the Brown-Rho scaling mechanism
\cite{dongbrs13,brown91,hatsuda,brown96,brown02,brown04} 
or the inclusion of three-nucleon potential $V_{3N}$ 
\cite{bogner05,hebeler11,holt12,coraggio14}. 
(The symbol $V_{NN}$ will be used to denote the two-nucleon
potential.)

\begin{figure}[here]
\begin{center}
%\scalebox{0.36}{\includegraphics[angle=-90]{11113newbr.eps3sym2015n0}}
\scalebox{0.36}{\includegraphics[angle=-90]{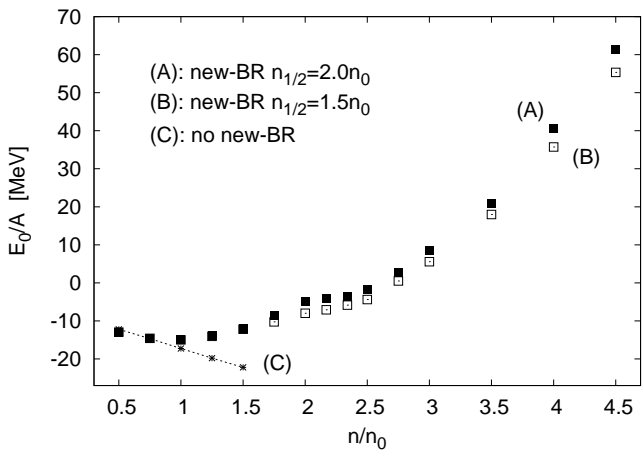}}
%figureeps
\caption{Comparison of the EoS for symmetric nuclear matter calculated with
and without the new-BR scaling. Transition densities of
$n_{1/2}=2.0n_0$  (solid square) and $1.5n_0$ (open square) are employed.
 See text for more explanations.}
\end{center}
\end{figure}

Realistic nucleon-nucleon potentials are mediated by the exchange
of mesons such as the $\pi$, $\rho$, $\omega$ and $\sigma$ mesons.
In constructing these potentials the meson-nucleon coupling constants 
(and in some cases their masses)
are adjusted to fit the `free-space' NN scattering data. 
Mesons in a nuclear medium, however,
can have properties (masses and couplings) that are 
different from those in free space, as the former
are `dressed' or `renormalized' by their interactions with the medium.
Thus, the NN potential in a medium of density $n$, 
usually denoted by $V_{NN}(n)$, should be
different from that in free space. 

 The well-known Brown-Rho (BR) scaling mechanism 
\cite{brown91,hatsuda,brown96,brown02,brown04} 
provides a schematic framework for deriving such density-dependent
interactions.
A main result is the universal scaling relation for nucleon and meson masses:
\begin{eqnarray}
&& \frac{m^*_{N}}{m_{N}}\simeq
\frac{m^*_{\sigma}}{m_{\sigma}}\simeq
\frac{m^*_{\rho}}{m_{\rho}}\simeq
\frac{m^*_{\omega}}{m_{\omega}}\simeq \Phi_{BR} (n), \nonumber \\
&& \Phi_{BR} (n)= 1-C\frac{n}{n_0},
\end{eqnarray}
where $m^*$ and $m$ denote respectively the in-medium and in-vacuum
masses, and the parameter $C$ has the value $0.15-0.20$. 
This scaling naturally
renders $V_{NN}$ a density-dependent interaction $V_{NN}(n)$.
As we shall describe later, it is remarkable 
that the above simple scaling law, derived in the context 
of chiral symmetry restoration in dense matter, 
would have important consequences for traditional nuclear 
structure physics. 

We have carried out several studies on the effects
of BR scaling on finite nuclei, nuclear matter and neutron stars
\cite{siu09,dong09,holtc1408,holtbrs07,dong11}.
Let us just briefly describe a few of them. For convenience in
implementing BR scaling, we have employed the BonnA and/or BonnS
one-boson-exchange potentials \cite{mach89,bonns} whose
parameters for $\rho$, $\omega$ and $\sigma$ mesons are 
scaled with the density
(in our calculations the meson masses and cut-off 
parameters are equally scaled). 
Note that $\pi$ is protected by chiral symmetry and is not scaled.
That we scale $\rho$ but not $\pi$ has an important consequence for the
tensor force, which plays an important role 
in the famous Gamow-Teller (GT) matrix element for the 
$^{14}C\rightarrow$ $^{14}N$ $\beta$-decay
\cite{holtc1408}. The tensor force from 
$\pi$- and $\rho$-meson exchange are of opposite
signs. A lowering of only $m_{\rho}$, 
but not $m_{\pi}$, can significantly
suppress the net tensor force strength and 
thus largely diminish the GT matrix element.
In addition, the scaling of the $\omega$ meson 
introduces additional short-distance
repulsion into the nucleon-nucleon interaction, which was 
found to also contribute to
the suppression \cite{holtc1409}. With BR-scaling we were 
able to satisfactorily account for the anomalously-long
$T_{1/2} \sim 5730$-yr lifetime of this decay \cite{holtc1408}.

It may be noted that the BR scaling as given in Eq.\ (65)
is a linear scaling, and it may be suitable 
  for low densities near $n_0$ only  
($\Phi _{BR}(n)$ of Eq.\ (67) may become negative for
large $n$). For high densities, we need a different scaling 
 such as the new-BR scaling \cite{dongnewbr13}. Before describing
this scaling, let us first describe briefly a 
$V_{low-k}$ ring-diagram formalism
\cite{siu09,dong09,dong11,dongnewbr13} 
 on which
our nuclear-matter calculations with the new-BR scaling are based. 

In this ring-diagram formalism 
\cite{siu09,dong09,dong11,dongnewbr13} 
the nuclear-matter ground-state energy is given by
the all-order sum of
the particle-particle hole-hole ($pphh$) ring diagrams
as shown in Fig.\ 3. 
 The ground-state energy of asymmetric nuclear matter is expressed as
$E(n,\alpha)=E^{free}(n,\alpha)+ \Delta E(n,\alpha)$
where $E^{free}$ denotes the energy for the non-interacting system
and $\Delta E$ is the energy shift due to the NN interaction.
 We include in general three types of ring diagrams, the
proton-proton, neutron-neutron and proton-neutron ones.
 The proton and neutron
Fermi momenta are, respectively,
$k_{Fp}=(3\pi^2n_p)^{1/3}$
and $k_{Fn}=(3\pi^2n_n)^{1/3}$, where $n_p$ and $n_n$ denote respectively
the proton and neutron density. The isospin asymmetry parameter is
$\alpha \equiv (n_n-n_p)/(n_n+n_p)$.
With such ring diagrams summed to all orders, 
 we have
\br\label{eng}
\Delta E(n,\alpha)&=&\int_0^1 d\lambda
\sum_m \sum_{ijkl<\Lambda}Y_m(ij,\lambda)  Y_m^*(kl,\lambda) \nonumber \\ 
&& \times \langle ij|V_{\rm low-k}|kl \rangle,
\er
where  $\lambda$ is a strength parameter,
integrated from 0 to 1. 
The transition amplitudes $Y$ are obtained from 
a $pphh$ RPA equation
\cite{siu09,dong09,dongbrs13}.
These $Y$ amplitudes represent
\br
&&Y_m(\alpha,\beta)=\langle\Psi _0|a_{\alpha}^{\dagger}
a_{\beta}^{\dagger}|\Psi _m\rangle, \nonumber \\
&& ~(\alpha,\beta)=(h,h')~or~(p,p'),
\er
where $\Psi _0$ and $\Psi _m$ are respectively the ground state
and $m$th excited state of nuclear matter with two additional
particles or holes.
$Y_m(p,p')$ is a measure of the particle-particle excitations in
the ground state. Setting $Y_m(p,p')$=0 means the ground state
is taken to be the  closed Fermi sea, and 
the above ring-diagram method reduces to the
usual HF method where only the first-order ring diagram 
(diagram (a) of Fig. 3)
is included for the energy shift. In this case, the 
above energy shift becomes
$\Delta E(n,\alpha)_{HF}=\frac{1}{2}
\sum n_i n_j\langle ij|V_{\rm low-k}|ij \rangle$ where
$n_k$=(1,0) if $k(\leq,>)k_{Fp}$ for  protons
and $n_k$=(1,0) if $k(\leq,>)k_{Fn}$ for  neutrons.

 The above $V_{low-k}$ ring-diagram framework has been applied to
symmetric and asymmetric nuclear matter \cite{siu09,dong09}
 and to the nuclear symmetry energy
\cite{dong11}.  This framework has also been tested by  
applying it  to dilute cold neutron matter in the  
limit that the $^1S_0$ scattering
length of the underlying interaction approaches infinity
\cite{siu08,dong10}. This limit 
-- which is a conformal fixed point --  is usually referred 
to as the unitary limit.
 For many-body systems at this limit, the ratio
$\xi \equiv E_0/E_0^{free}$ is
expected to be a universal constant of value $\sim 0.44$. 
($E_0$ and $E_0^{free}$
are, respectively, the interacting and 
non-interacting ground-state
energies of the many-body system.)  The above ring-diagram 
method has been
used to calculate
neutron matter using several very different unitarity potentials
(a unitarity CDBonn potential  obtained by tuning
its meson parameters,
 and several square-well unitarity potentials) \cite{siu08,dong10}.
The $\xi$ ratios
given by our calculations for all these different unitarity potentials are
all close to 0.44, in good agreement with the Quantum-Monte-Carlo results
(see \cite{dong10} and references quoted therein).
In fact our ring-diagram results for $\xi$ are significantly better than
those given by HF and BHF (Brueckner HF) \cite{siu08,dong10}. 
The above unitary calculations have provided
satisfactory results,  supporting the reliability of
 our $V_{low-k}$ ring-diagram framework for 
calculating the nuclear matter EoS.

We now describe the new-BR scaling \cite{dongnewbr13} 
on which the results shown in Fig.\ 15 are based.
The idea behind this scaling is that when a large number of
skyrmions as baryons are put on an FCC (face-centered-cubic) crystal
to simulate dense
matter, the skyrmion matter undergoes a transition to a matter consisting
of half-skyrmions \cite{goldhaber} in BCC (body-centered-cubic) 
configuration at a density
that we shall denote as $n_{1/2}$. This density is difficult to pin
down precisely but it is more or less independent of the mass of the
dilaton scalar, the only low-energy degree of freedom that is not
well-known in free space. The density at which this 
occurs has been estimated to lie typically
between 1.3 and 2 times normal nuclear matter density $n_0$
~\cite{half}.
In our model, nuclear matter is separated into two regions I and II
respectively for densities $n\leq n_{1/2}$ and $n>n_{1/2}$. As inferred
by our model, they have different scaling functions
\begin{equation}
\Phi_i (n) = \frac{1}{1+C_i \frac{n}{n_0}},~~ i=I,II.
\end{equation}
%The above region-II scaling is the new-BR scaling mentioned earlier. 

It has been found that both the BR and new-BR scalings are important
 for nuclear-matter saturation
\cite{siu09,dong09,holtbrs07,dongnewbr13}. The EoS shown by (A) and
(B) of Fig.\ 15
 are obtained with the new-BR scaling with
 $n_{1/2} =1.5 n_0$ and 2.0$n_0$ respectively.
 Both give an energy per nucleon 
$E_0/A=-15$ MeV, saturation density $k_F=1.30 fm^{-1}$
 and compression modulus $K$=208 MeV, all in satisfactory
agreement with the empirical values \cite{dongnewbr13}.
We believe that this scaling works 
well for low densities of $n < \sim 1.5n_0$.

 As described in
\cite{dongnewbr13}, we employ in our new-BR scaling calculations
 the BonnS potential \cite{bonns} with scaling parameters
$C_{\rho}$=0.13,
$C_{\sigma}$=0.121,
$C_{\omega}$=0.139 ,
$C_{N}$=0.13 and
$C_{g,\rho}=C_{g,\omega}$=0 for region I, where $C_g$ is the associated
scaling of the meson-nucleon coupling constant.
For region II the scaling parameters are
$C_{\rho}$=0.13,
$C_{\sigma}$=0.121,
$C_{\omega}$=0.139 ,
$C_{g,\rho}$=0.13, $C_{g,\omega}$=0 and $m^*_N/m_N=y(n)$=0.77 for (A)
and 0.78 for (B). 
Note that this scaling has some special features:
In region I the coupling constants $g_{\rho N}$ and $g_{\omega N}$
are not scaled, while in region II only the coupling constant $g_{\rho N}$
is scaled. Also in region II 
the nucleon mass is a density-independent 
constant as given earlier. 
Note that our choices
for the $C$ parameters are consistent with the Ericson scaling
which is based on a scaling relation for the quark condensate
$\frac{<\bar qq>^*}{<\bar qq>}$ \cite{ericson}. According to this
scaling, at low densities one should have $C\simeq D/3$ with
$D=0.35 \pm 0.06$.

%\begin{figure}[here]     
%\scalebox{0.42}{\includegraphics[angle=-90]{818withskyrmesym}}
%\scalebox{0.42}{\includegraphics[angle=-90]{10withskyrmesym}}
%\caption{Ring-diagram EOS for symmetric nuclear matter with the interaction
%being the sum of  $V_{low-k}$  and the Skyrme density dependent
% force of Eq. (66). 
%Four sets of results are shown for CDBonn and BonnA potentials 
%with $\Lambda$=3 and
% 3.5 $fm^{-1}$. A common Skyrme force of $t_3$=2000 MeV-$fm^6$ 
%and $x_3=0$ is 
%employed.}
%\end{figure}          

\begin{figure}[t]
\begin{center}
%\scalebox{0.32}{\includegraphics[angle=-90]{91012newbr.epsneu7778press}}
%\scalebox{0.32}{\includegraphics[angle=-90]{91012newbr.epssym7778press}}
\scalebox{0.32}{\includegraphics[angle=-90]{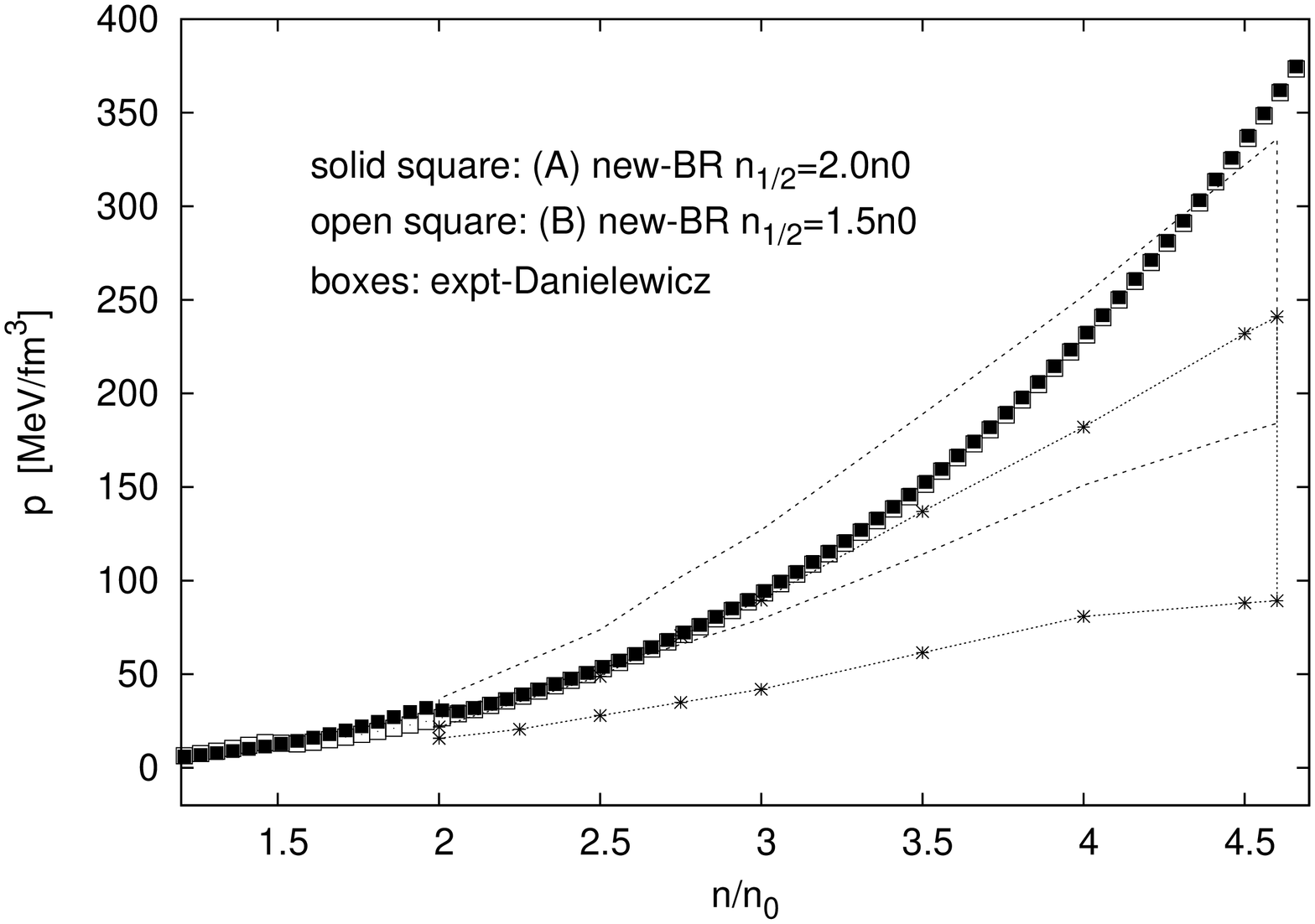}}
\scalebox{0.32}{\includegraphics[angle=-90]{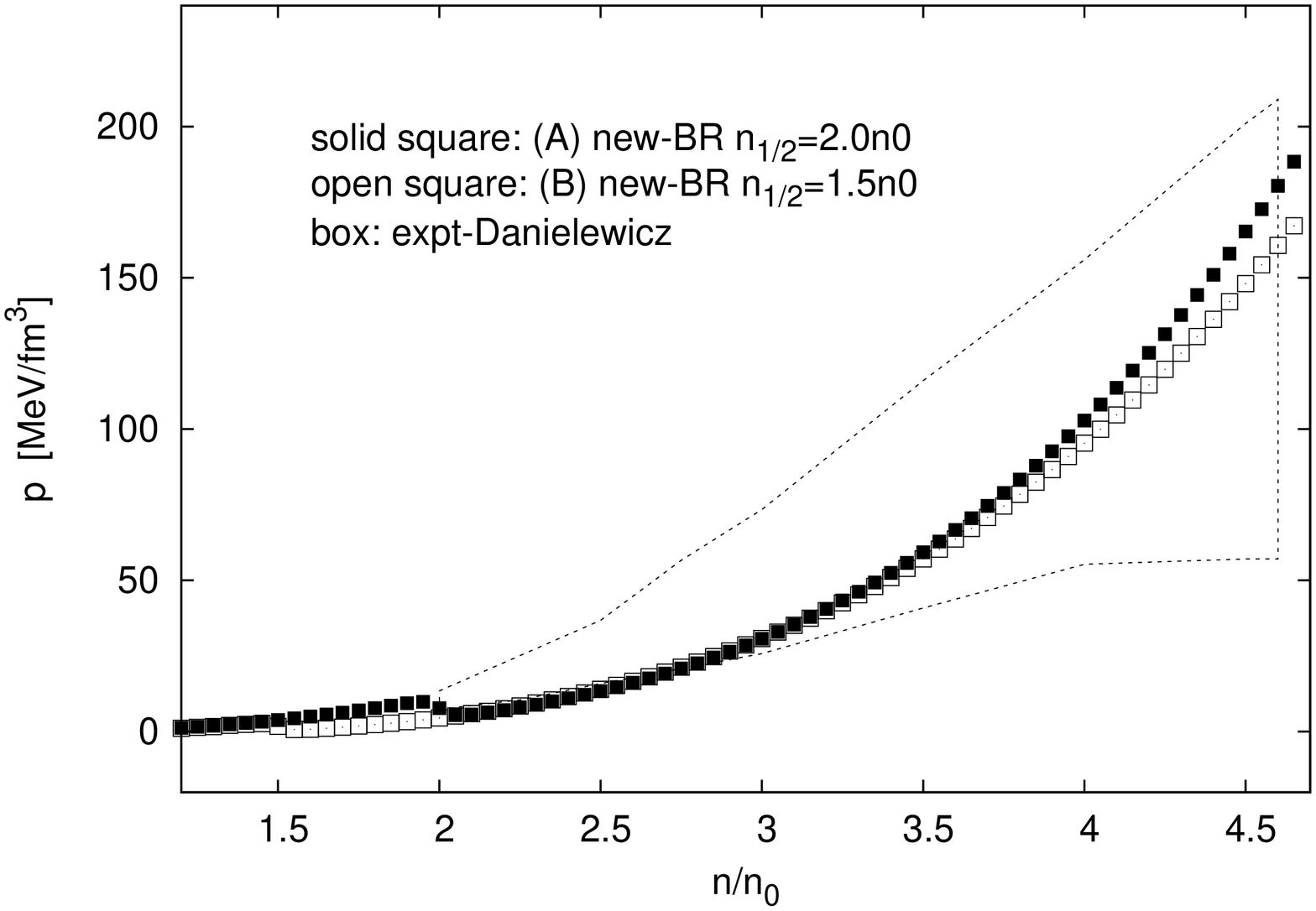}}
%figureeps
\caption{Comparison of our calculated $p(n)$ for 
neutron matter (upper panel)
and symmetric  nuclear matter (lower panel) 
with the empirical 
constraints of Danielewicz  et al.\ \cite{daniel02}.
For neutron matter, the stiff and soft constraints are
denoted respectively by the upper and lower box.}
\end{center}
\end{figure}

>From heavy-ion collision experiments, 
Danielewicz {\it et al.} \cite{daniel02} have
obtained constraints for the pressure-density EoS $p(n)$
of nuclear matter up to densities $\sim 4.5 n_0$.
To  test this new-BR scaling in the high-density region,
 the pressure EoS $p(n)$ up to the above densities
has been calculated.
 \cite{dongnewbr13}
The results are presented in Fig.\ 16 where the 
calculated $p(n)$ for neutron matter (upper panel)
and symmetric nuclear matter (lower panel) are compared with the
Danielewicz constraints \cite{daniel02}. For neutron matter,
there are two constraints, one for stiff EoS (upper box)
and the other for soft EoS (lower box).
 The EoSs calculated with parameters A and B are denoted by
 solid- and open-square respectively. 
As seen, the results are in satisfactory
agreements with the empirical constraints of \cite{daniel02}.

 Also based on such experiments,
there has been much progress  in determining
the nuclear symmetry energy $E_{sym}$ up to densities as high
as $\sim 5n_0$
\cite{li05,li08,tsang09}.
Thus an application of the new-BR  scaling
to the calculation of $E_{sym}$ would provide an important test
for this scaling in the region with $n>n_{1/2}$.
As displayed in Fig.\ 17,  the symmetry
energies calculated from the new-BR EoS \cite{dongnewbr13} 
are in good agreement  with 
the experimental constraints of Li et al. \cite{li05,li08} 
and Tsang et al. \cite{tsang09}. 
As seen, there are two constraints from Li, one for stiff EoS and the
other for soft EoS. The calculated results are close to the stiff constraint
at high denities, but slightly lower than both at low densities.

The EoS at high densities ($n \simeq 5-10n_0$) is important for
neutron-star properties. Thus an application of the new-BR scaling
to neutron star structure would provide a further test. 
As shown in Fig.\ 18,
the maximum mass of pure-neutron stars  calculated from
the new-BR EoS  is about  $2.4 M_{\odot}$,
slightly larger than the maximum mass $M \sim 2M_{\odot}$
observed in nature   
\cite{dongnewbr13,demorest2010,antoniadis13}.
The central
density  of the neutron star is $\sim 5n_0$. At  densities
as high as this central density,
 how to scale the hadrons
with the medium is an interesting  question and should
be further studied.
 We note that the above scaling is only `inferred'  by the
Skyrmion-half-Skyrmion model \cite{dongnewbr13}.
Although the initial results obtained with this model
are promising,  further studies of this model should be
useful and of interest.

\begin{figure}[t]
\begin{center}
%\scalebox{0.34}{\includegraphics[angle=-90]{11113newbr.eps4esym2015}}
\scalebox{0.34}{\includegraphics[angle=-90]{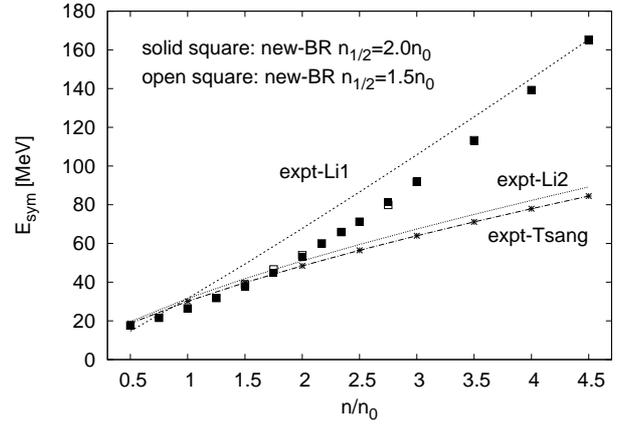}}
%figureeps
\caption{
Comparison of our calculated
nuclear symmetry energies
with the empirical upper (expt-Li1)
and lower (expt-Li2) constraints
of Li et al. \cite{li05} and the empirical results of
Tsang et al. (expt-Tsang) \cite{tsang09}.
}
\end{center}
\end{figure}
\begin{figure}[t]
\begin{center}
%\scalebox{0.33}{\includegraphics[angle=-90]{81512newbr.epsnstar1520}}
\scalebox{0.33}{\includegraphics[angle=-90]{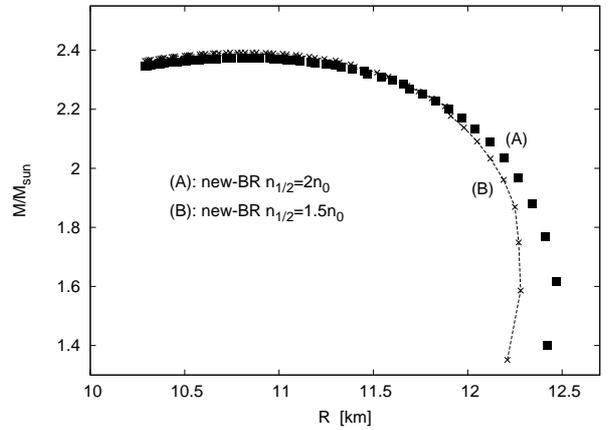}}
%figureeps
\caption{Mass-radius relations of
neutron stars calculated
with new-BR scalings using $n_{1/2}$ = 2.0 (A) and 1.5$n_0$ (B).
The maximum neutron-star
mass and its radius for these two cases are
respectively \{$M_{\rm max} = 2.39 M_{\odot}$ and $R = 10.90$ km\} and
\{ $M_{\rm max} = 2.38 M_{\odot}$ and $R= 10.89$ km\}. }
\end{center}
\end{figure}

  For densities near $n_0$, the BR scaling of Eq.\ (67)
and new-BR scaling of Eq.\ (70) are practically the same,
 both rendering the NN interaction into a density-dependent 
 one. This density dependence has played
an important role in describing nuclear properties
such as nuclear-matter saturation. \cite{kuo14}
It is of interest  that this importance was 
already recognized in the well-known
 Skyrme effective interaction
\cite{skyrme} of the form
\be
V_{skyrme}=V_{sky}(\vec r_1 -\vec r_2) 
+ t_3 \delta (\vec r_1 - \vec r_2 ) \delta (\vec r_2 -\vec r_3),
\ee
where $V_{sky}$ is a zero-range two-body interaction.
The second term is a zero-range three-body interaction,
and by integrating out one participating nucleon
over the Fermi sea it becomes a density-dependent two-body
force $D_{sky}$, namely
\begin{eqnarray}
V_{skyrme}&=&V_{sky}(\vec r_1-\vec r_2)+D_{sky}(\vec r_1-\vec r_2), \nonumber \\ 
D_{sky}&=&\frac{1}{6}(1+x_3P_{\sigma})t_3\delta (\vec r_1 - \vec r_2)
n(\vec r_{av}),
\end{eqnarray}
where  
 $\vec r_{av}\equiv (\vec r_1 +\vec r_2)/2$ and both
$t_3$ and $x_3$ are strength parameters. In calculations
with Skyrme interactions, the inclusion of $D_{sky}$ is
essential for nuclear matter saturation. 

%Note that the new-BR EoSs are different in  regions 
%below and above the transition density
%$n_{1/2}$. To have a better agreement between the two calculations,
%we may need to use different $D_{sky}$ for the two regions. 

As observed in Ref.\ \cite{dong09} the combined
 potential given by the sum of the
unscaled-$V_{NN}$ and $D_{sky}$,  the Skyrme 
density-dependent interaction
shown above, can  give equally satisfactory nuclear matter
saturation properties as the BR-scaled $V_{NN}$. 
(We use $V_{NN}$ to denote the two-nucleon interaction.)
Similar equivalence is also noted for the new-BR scaling,
as illustrated
in Fig.\ 19. There a qualitative agreement
is seen up to densities  $\sim 4n_0$
  between
 the EoS for symmetric nuclear
matter calculated with a new-BR-scaled $V_{NN}$ and that
with a combined interaction ($V_{NN}+D_{sky}$).
(The parameters of $t_3=5000 MeV~fm^6$ and $x_3$=0
are used for the $D_{sky}$ calculation of Fig.\ 19.)

We now come to the three-nucleon force $V_{3N}$.
As we have just discussed,  nuclear matter calculations with
either BR or  new-BR scaled
 $V_{NN}$ can satisfactorily describe nuclear matter saturation
properties but not so with the unscaled $V_{NN}$. 
It is well known that  there are many-nucleon forces $V_{3N}$ and beyond.
Especially for nucleons in dense medium, the effects of such many-body
forces are likely to be important.
Can the calculations using $V_{NN}$ (unscaled) plus $V_{3N}$ also
give satisfactory nuclear saturation properties?

\begin{figure}[t]
\begin{center}
%\scalebox{0.33}{\includegraphics[angle=-90]{90815newbr2n0t3b5000.epssym}}
%\scalebox{0.33}{\includegraphics[angle=-90]{915v3n414t3b5000.epssym}}
\scalebox{0.33}{\includegraphics[angle=-90]{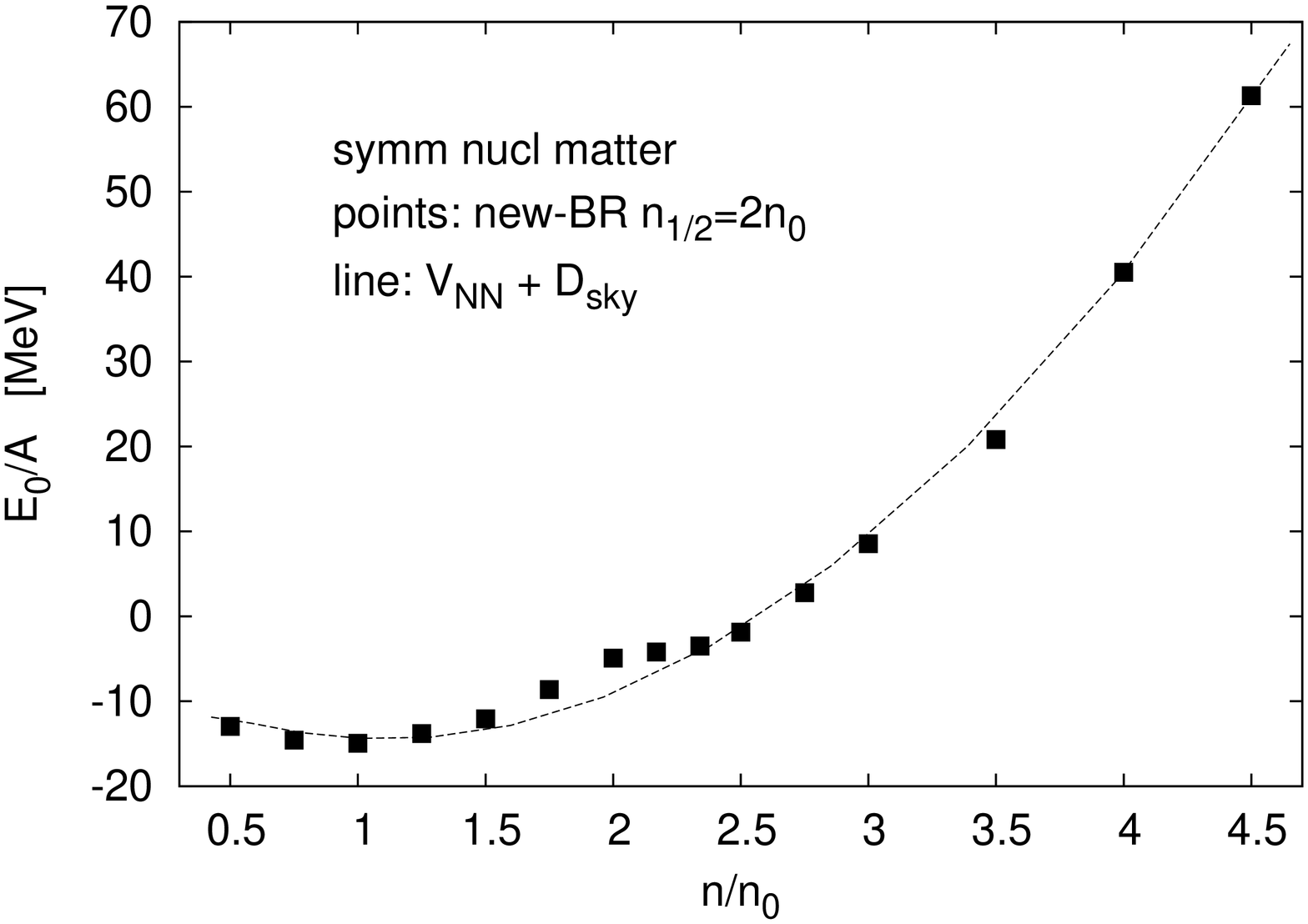}}
\scalebox{0.33}{\includegraphics[angle=-90]{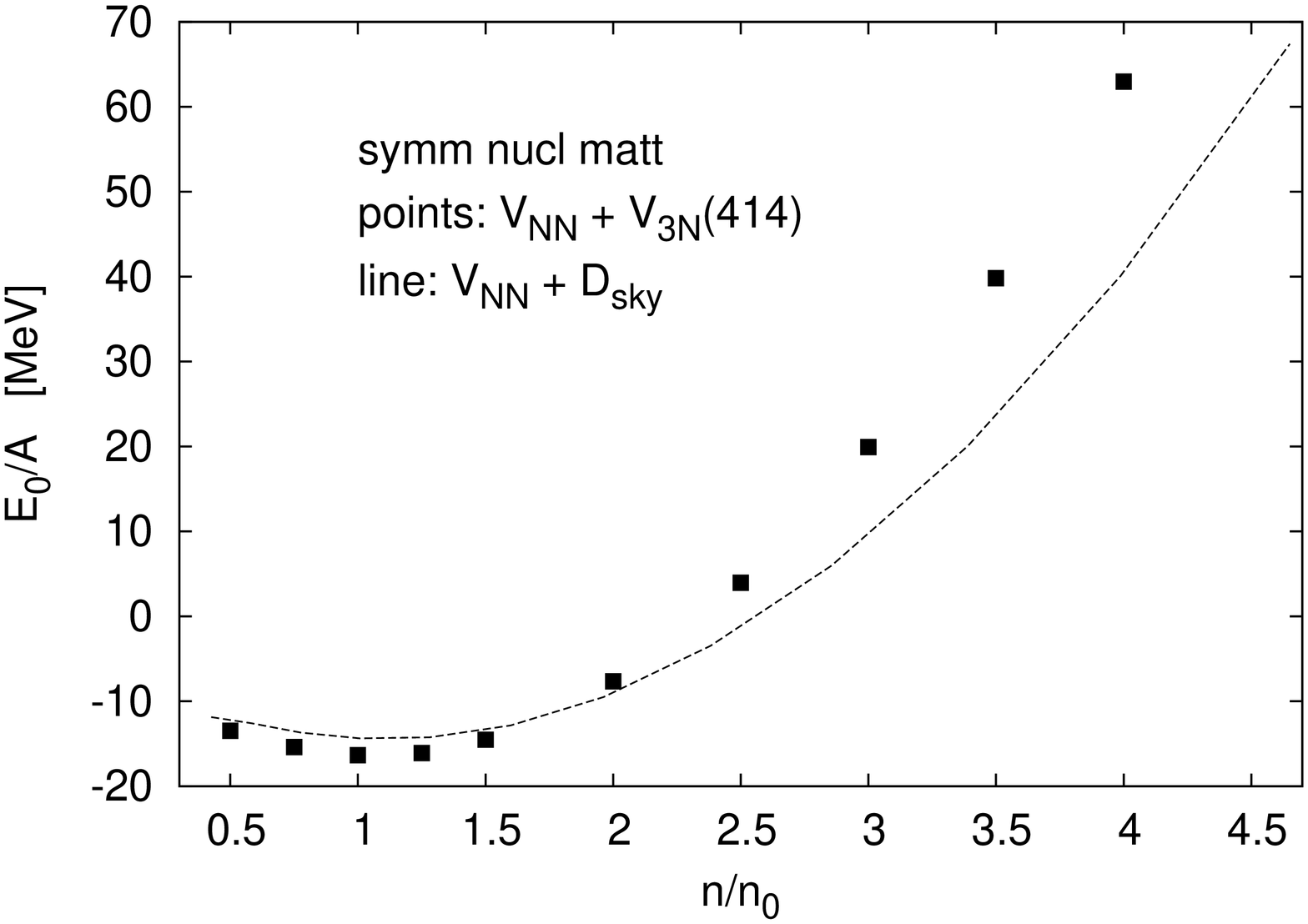}}
%figureeps
\caption{Comparison of ring-diagram EoS for symmetric nuclear matter 
calculated with new-BR-scaled $V_{NN}$ (upper panel), 
$V_{NN}$+$\bar V_{3N}(414)$ (lower panel),
and $V_{NN}+D_{sky}$. See text for further explanations.}
\end{center}
\end{figure}
 
In the following we shall study this question.
 The lowest-order (NNLO) chiral three-nucleon interaction 
$V_{3N}$ will be considered. 
This interaction is  of the form
 $V_{3N} = V_{3N}^{2\pi} + V_{3N}^{1\pi} + V_{3N}^{ct}$ where 
\begin{equation}
V_{3N}^{(2\pi)} = \sum_{i\neq j\neq k} \frac{g_A^2}{8f_\pi^4}
\frac{\vec{\sigma}_i \cdot \vec{q}_i \, \vec{\sigma}_j \cdot
\vec{q}_j}{(\vec{q_i}^2 + m_\pi^2)(\vec{q_j}^2+m_\pi^2)}
F_{ijk}^{\alpha \beta}\tau_i^\alpha \tau_j^\beta, \nonumber
\label{3n1}
\end{equation}
\begin{equation}
V_{3N}^{(1\pi)} = -\sum_{i\neq j\neq k}
\frac{g_A c_D}{8f_\pi^4 \Lambda_\chi} \frac{\vec{\sigma}_j \cdot
\vec{q}_j}{\vec{q_j}^2+m_\pi^2}
\vec{\sigma}_i \cdot
\vec{q}_j \, {\vec \tau}_i \cdot {\vec \tau}_j , \nonumber
\label{3n2}
\end{equation}
\begin{equation}
V_{3N}^{(\rm ct)} = \sum_{i\neq j\neq k} \frac{c_E}{2f_\pi^4 \Lambda_\chi}
{\vec \tau}_i \cdot {\vec \tau}_j,
\label{3n3}
\end{equation}
where $g_A=1.29$, $f_\pi = 92.4$ MeV, $\Lambda_{\chi} = 700$ MeV,
$m_{\pi} = 138.04$ MeV,
$\vec{q}_i=\vec{p_i}^\prime -\vec{p}_i$ is the difference between the
final and initial momentum of nucleon $i$ and
\br
 F_{ijk}^{\alpha \beta} &=& \delta^{\alpha \beta}\left (-4c_1m_\pi^2
 + 2c_3 \vec{q}_i \cdot \vec{q}_j \right )
\nonumber \\
&& + c_4 \epsilon^{\alpha \beta \gamma} \tau_k^\gamma \vec{\sigma}_k
\cdot \left ( \vec{q}_i \times \vec{q}_j \right ). 
\label{3n4}
\er
The parameters $c_1, c_3, c_4$ are constrained by $\pi N$ scattering and
can then be fitted within these uncertainties to peripheral nucleon-nucleon
scattering phase shifts \cite{entem03}. The parameters $c_D$ and $c_E$
must be determined by three-nucleon observables.
Navratil et al.\ \cite{navratil07} fit to the binding energies of 
$^3H$ and $^3He$, which determined a curve of consistent $c_D$ and $c_E$ 
values. To uniquely determine  $c_D$ and $c_E$, 
one can fit also the $\beta$-decay lifetime of $^3H$ 
\cite{gardestig06,gazit08,marcucci12}.
Recently Coraggio et al.\ \cite{coraggio13,coraggio14}
have carried out extensive nuclear matter calculations using
a chiral $V_{NN}$ \cite{chiralvnn} and the above $V_{3N}$ with
its $c_D$ and $c_E$ parameters determined also from the binding energies
of three-nucleon systems and the lifetime of $^3H$.

The values of the ($c_D$, $c_E$) parameters so determined are $(-0.40, -0.07)$,
$(-0.24, -0.11)$ and $(0.0, -0.18)$ respectively for variations in the 
resolution scale defined by the momentum-space cutoff 
$\Lambda=(414,450,500)$ MeV \cite{coraggio14}, where $\Lambda$ enters
the regulating function
\be
f(p',p)=exp[-(p'/\Lambda)^{2n}-(p/\Lambda)^{2n}]
\label{regfunc}
\ee 
that multiplies the nucleon-nucleon potential. In Eq.~(\ref{regfunc}) 
$p'$ and $p$ denote the relative momenta of the incoming and
outgoing nucleons, and the regulator exponent $n$ for the above 
three values of $\Lambda$ are
$n = (10,3,2)$ respectively.
As shown above, the $c_D$ and  $c_E$ parameters for 
$\Lambda= (414, 450, 500)$ MeV do not differ largely from each other.
We shall denote the two- and three-nucleon potentials corresponding
to a chosen $\Lambda$ as $V_{NN}(\Lambda)$ and $V_{3N}(\Lambda)$
respectively. In Refs.\ \cite{coraggio13,coraggio14} it was shown that
low-momentum chiral potentials at the resolution scales $\Lambda = 414$ 
and 450 MeV have perturbative properties similar to the 
renormalization-group evolved potential $V_{low-k}$.
In the present work, we have employed only one of them,
the $\Lambda$=414 set.
We plan to carry out calculations using the other two $\Lambda$
choices in a future work. 

 By integrating out one participating nucleon over the Fermi sea,
 Holt, Kaiser and Weise \cite{holtc1409,holt10} have reduced $V_{3N}$ 
 to a density-dependent two-body force $\bar V_{3N}$.
Comparing with $V_{3N}$, $\bar V_{3N}$  is much more
convenient for nuclear many-body calculations. Briefly speaking,
they are related  by
\begin{eqnarray}
 &&V_{3N}= \frac{1}{36} \sum \langle 123  |V_{3N}|456 \rangle
 a_3^+a_2^+a_1^+a_4a_5a_6, \nonumber \\  
 &&\bar V_{3N}= \frac{1}{4} \sum \langle 12  | D_{2N}|45 \rangle
~ a_2^+a_1^+a_4a_5, \nonumber \\
&&\langle ab|  D_{2N}| cd  \rangle
= \sum _{h\leq k_F} \langle abh| V_{3N}|cdh   \rangle .
\end{eqnarray} 
It is seen that 
 $\bar V_{3N}$ is a density ($k_F$) dependent two-body interaction.
The basis states $|456 \rangle$, $|45 \rangle$, $\cdots$ in Eq.\ (78)
 are all
anti-symmetrized and normalized.
Note as well that there is a $n$-body counting factor $C(n)$
to be included in calculations:
we have  $C(n)$ = (1, 1/2, and 1/3) respectively for (2-, 1-, 0-) body
vertices. Thus the vertex in (a) of Fig.\ 3
 is ($V_{NN}+\bar V_{3N}/3$), and each vertex  in (b) and (c) 
 is ($V_{NN}+\bar V_{3N}$). In the 1-body
 diagram $d1$ of Fig.\ 4 the vertex is $(V_{NN}+ \bar V_{3N}/2)$.

 As an initial application, we have performed ring-diagram
calculations of nuclear matter EoS and nuclear symmetry
energy using the combined potential $V_{NN}(414)$ plus
$\bar V_{3N}$ derived
from $V_{3N}(414)$. The  ring-diagram method used here
is the same as used earlier for the new-BR calculations. 
The resulting EoS for symmetric nuclear matter is
presented in Fig.\ 19 (lower panel), denoted by 
`$V_{NN}+V_{3N}(414)$'. This EoS has
ground-state energy per nucleon $E_0/A=-16.4$ MeV, compression modulus
$K=198$ MeV and saturation density $k_F=1.36 fm^{-1}$,
all in satisfactory agreement with the 
empirical values. Our EoS calculated with $V_{3N}(414)$ has
worked well in the low-density region near $n_0$.
In this region, this EoS and the
new-BR one shown in the upper panel of the figure
agree well with each other, indicating that 
the new-BR-scaled $V_{NN}$ and the combined potential
$(V_{NN}(414)+V_{3N}(414))$ can both describe nuclear matter
at low density near $n_0$ satisfactorily.
But for densities beyond $\sim 2n_0$, the EoS given by these
two potentials have significant differences as shown by Fig.\ 19.
 In the lower panel of Fig.\ 19, we also present the EoS given by
the combined potential
 ($V_{NN}(414)+D_{sky}$) where $D_{sky}$ is the same Skyrme-type
density-dependent interaction used in the
upper panel of the figure.
The same ring-diagram method
is used for both EoS. 
As seen, this EoS and the one  from
 ($V_{NN}(414)+V_{3N}(414)$) 
  are in good
agreement for densities lower than $\sim 2 n_0$, indicating
that for such low densities the effect of $V_{3N}$ can be
well reproduced by $D_{sky}$. But   at higher densities,
the $D_{sky}$ EoS is significantly lower. To have better agreement 
between the two, we may need to use
 a $D_{sky}$ with different
$t_3$ and $x_3$ parameters.  This possibility is being studied by us and
some recent work in this direction can 
be found in Refs.\ \cite{brown14,rrapaj15}.

Medium modifications to the nucleon-nucleon interaction resulting
from Brown-Rho scaling or three-nucleon forces have also succeeded in 
explaining the very long lifetime ($T_{1/2} \sim 5730$\,yrs)
of the $^{14}C\rightarrow$ $^{14}N$ $\beta$-decay. The expected
lifetime is only several hours, and in order to 
achieve the empirical lifetime 
a precise cancellation in the Gamow-Teller
matrix element to the order of one part in a thousand is required.
Calculations using $V_{NN}$ (unscaled) find only a small suppression \cite{aroua03}
of the transition, 
and it is only with the inclusion of medium modifications that it has been 
possible to explain the long lifetime \cite{holtc1408,holtc1409,maris11}.

  Based on the above 
 $(V_{NN}(414)+V_{3N}(414))$ EoS we have also calculated
the nuclear symmetry energy $E_{sym}$ using a method
as described in Ref.\ \cite{dong11}.
Lattimer and Lim \cite{lattimer13} have studied the constraints
on $E_{sym}$ and $L$ (defined as $3u(dE_{sym}/du),~u\equiv n/n_0$).
At density $n=n_0$ their constraints are
$29.0 \leq E_{sym}/MeV \leq 32.7$ and
$40.5 \leq L/MeV \leq 61.9$. At $n=n_0$, 
our results  are
$E_{sym}= 31.06$ MeV and $L= 47$ MeV, both in good agreement
with the Lattimer-Lim constraints.
At densities higher than $\sim 2n_0$, our calculated $E_{sym}$
exhibits, however, a supersoft behavior (namely it decreases
with density after reaching a maximum near $2n_0$). 
As discussed in Refs.\ \cite{li05,li08,brown00,xiao09,li10}, 
predictions of  $E_{sym}$ at high densities are
rather diverse, depending on the nucleon-nucleon
interactions employed. Further studies of $E_{sym}$
at densities beyond $\sim 2n_0$ are certainly needed and of interest.

\section{Summary and discussion}
 As illustrated by the pioneering work of Talmi \cite{talmi},
nuclear shell-model calculations in small model spaces can describe
empirical nuclear properties highly successfully.
An important ingredient in such calculations is the
use of empirically determined effective interactions.
It is of interest to see if such interactions can 
be derived microscopically
from an underlying NN interaction and how. 
This question has been widely
studied in past years. In this paper 
several developments related to this question have been discussed.

First we reviewed the $\hat Q$-box folded-diagram theory,
with which the full-space nuclear Hamiltonian can be reduced to
a model-space effective one
of the form $PH_{eff}P=P(H_0^{expt}+V_{eff})P$.
Here $H_0^{expt}$ is the sp Hamiltonian extracted from
 experimental sp energies, 
and the effective interaction  is given by a
folded-diagram expansion in terms of the irreducible vertex
function $\hat Q$-box. Methods for summing up this series
for both degenerate and non-degenerate model spaces
are discussed. To calculate $V_{eff}$, a first step
is to calculate the $\hat Q$-box, which is composed
of irreducible valence-linked diagrams, from the input
NN interactions.

 We have reviewed the low-momentum interaction $V_{low-k}$
which has been commonly used in the $V_{eff}$ calculations.
This interaction has been obtained from realistic $V_{NN}$
by integrating out their high-momentum components
beyond a decimation scale $\Lambda$ using a renormalization-group
procedure. The resulting $V_{low-k}$ interaction
is a smooth potential (without strong repulsive core)
which is convenient for calculations. Furthermore
such interactions derived from different $V_{NN}$
models are nearly the same, for $\Lambda$ less than $\sim 2 fm^{-1}$,
leading to a nearly unique low-momentum interaction.
 As indicated by the few sample calculations presented
earlier,
shell-model calculations using the  effective  
interaction derived from $V_{low-k}$ have indeed worked rather
successfully in describing low-energy experimental data.

Turning to nuclear systems at densities near and above $n_0$,
a long standing problem has been that
 calculations with the two-nucleon potential $V_{NN}$ alone are unable
to give satisfactory saturation properties.  We have
demonstrated that this shortcoming
can be amended by including either the new-BR scaling or $V_{3N}(414)$. 
As illustrated in Fig.\ 19,  
the EoS
from new-BR ($n_{1/2}=2n_0$) and ($V_{NN}+V_{3N}(414)$) are close
 to each other for densities below $\sim 2 n_0$.
Both can reproduce the empirical nuclear saturation properties
well. In our EoS calculations we have employed a ring-diagram
method where the particle-particle hole-hole ring diagrams
are summed to all orders.
The symmetry energy at densities near $n_0$ can also be well
reproduced by the above two approaches. 
 There may be some underlying equivalence between 
new-BR scaling and $V_{3N}$ at low densities, 
and its further study will be interesting
as well as useful. 

A similar equivalence is observed concerning
the famous Gamow-Teller (GT) matrix element for the 
$^{14}C\rightarrow$ $^{14}N$ $\beta$-decay, which has an
anomalously long half-life of $\sim 5730$yrs. 
The use of the BR-scaled $V_{NN}$ has been crucial
in highlighting the important role played by medium effects not 
normally included in many-body perturbation theory calculations
with free-space two-body interactions. These medium effects can 
be recast in the language of three-body forces, which have been 
equally successful in describing the transition. 

But at densities higher than $\sim 2n_0$, 
the nuclear-matter EoS  given by 
these two approaches are significantly different,
as indicated by Fig.\ 19.
At such high densities, the symmetry energies given
by them are also very different. 
Studies of these differences and their impact on neutron star
properties, such as the mass-radius relationship and 
maximum mass, will be of interest.

 We have compared the  new-BR and $V_{3N}$ EoSs with 
the EoS given by (unscaled-$V_{NN}$ + $D_{sky}$). 
It is observed that all three are in good qualitative
agreement in the density region below $\sim 2 n_0$.
That the density-dependent effects
on symmetric nuclear matter near $n_0$  from the new-BR 
scaling and from $V_{3N}$
may be well reproduced by an empirical density-dependent
force of the Skyrme type is an interesting result, indicating
that the empirical
Skyrme force may have a microscopic connection 
with the new-BR scaling and/or $V_{3N}$. 

\vskip 0.3cm
{\bf Acknowledgement} 
We would like to acknowledge the inspiration Aage Bohr
and Ben Mottelson have been to our field over the decades,
as well as to us personally. Two of us (TTSK and EO)
cherish vivid memories from our visits to Copenhagen
during the golden years of nuclear physics.
We are grateful to Mannque Rho, Ruprecht Machleidt 
and Luigi Coraggio for many helpful discussions. 
This work was supported in part
by the Department of Energy under Grant No.\ DE-FG02-88ER40388
and DE-FG02-97ER-41014.

%We thank J. Dudek for inviting us to submit this review
%for publication in " Physica Scripta Special Edition commemorating
%the 40-year anniversary of the Nobel Prize for
%A. Bohr, B. Mottelson and L. Rainwater".  

\end{document}